\documentclass[aps,prb,twocolumn,superscriptaddress]{revtex4-1}
\usepackage{graphicx}
\usepackage{multirow}
\usepackage{color}
\usepackage{amsmath}
\usepackage{amsfonts}
\usepackage{amssymb}
\usepackage{graphicx}
\usepackage{epstopdf}
\usepackage{empheq}
\usepackage{float}
\usepackage{cases}
\usepackage{mathtools}
\usepackage{dcolumn}
\usepackage{bbold}
\usepackage{bm}

\DeclarePairedDelimiter\ket{\lvert}{\rangle}

\newcommand\zero{\mathbb{0}}
\newcommand\one{\mathbb{1}}

\renewcommand\vec[1]{\bm{#1}}

\definecolor{orange}{RGB}{255,127,0}

\newcommand{\CORR}{}
\newcommand{\CORRR}{}
\newcommand{\CORRRR}{}

\begin{document}

\title{Hole-phonon interactions in quantum dots: Effects of phonon confinement and encapsulation materials on spin-orbit qubits}

\author{Jing Li}
\email{jing.li@cea.fr}
\affiliation{Univ. Grenoble Alpes, CEA, IRIG-MEM-L\_Sim, F-38000, Grenoble, France}
\author{Benjamin Venitucci}
\email{benjamin.venitucci@cea.fr}
\affiliation{Univ. Grenoble Alpes, CEA, IRIG-MEM-L\_Sim, F-38000, Grenoble, France}
\author{Yann-Michel Niquet}
\email{yniquet@cea.fr}
\affiliation{Univ. Grenoble Alpes, CEA, IRIG-MEM-L\_Sim, F-38000, Grenoble, France}

\date{\today}

\begin{abstract}
Spin-phonon interactions are one of the mechanisms limiting the lifetime of spin qubits made in semiconductor quantum dots. At variance with other mechanisms such as charge noise, phonons are intrinsic to the device and can hardly be mitigated. They set, therefore fundamental limits to the relaxation time of the qubits. Here we introduce a general framework for the calculation of the spin (and charge) transition rates induced by bulk (3D) and strongly confined 1D or 2D phonons. We discuss the particular case of hole spin-orbit qubits described by the 6 bands $\vec{k}\cdot\vec{p}$ model. We next apply this theory to a hole qubit in a silicon-on-insulator device. We show that spin relaxation in this device is dominated by a band mixing term that couples the holes to transverse acoustic phonons through the valence band deformation potential $d$, and optimize the bias point and magnetic field orientation to maximize the number of Rabi oscillations $Q$ that can be achieved within on relaxation time $T_1$. Despite the strong spin-orbit coupling in the valence band, the phonon-limited $Q$ can reach a few tens of thousands. We next explore the effects of phonon confinement in 1D and 2D structures, and the impact of the encapsulation materials on the relaxation rates. We show that the spin lifetimes can depend on the structure of the device over micrometer-long length scales and that they improve when the materials around the qubit get harder. Phonon engineering in semiconductor qubits may therefore become relevant once the extrinsic sources of relaxation have been reduced.
\end{abstract}

\maketitle

\section{Introduction}

Spins in semiconductor quantum dots are considered as an attractive platform for quantum computing.\cite{Kane98,Loss98} Silicon\cite{Pla12,Zwanenburg13} has, in particular, garnered much interest as a host material for spin qubits because it can be isotopically purified in order to get rid of the nuclear spins that have undesirable hyperfine interactions with the electron spins. As a matter of fact, very long electron spin lifetimes $T_1$ have been measured on donors\cite{Tyryshkin12} and electrostatically defined quantum dots in silicon.\cite{Veldhorst14} This promoted the demonstration of high fidelity single and two qubit gates in this material.\cite{Kawakami14,Veldhorst15b,Takeda16,Yoneda18,Watson18,Zajac18,Huang19}

Among other specific features of silicon, intrinsic spin-orbit interactions are known to be weak in the conduction band.\cite{Zwanenburg13} This decouples the electron spins from electrical and charge noise as well as phonons and further enhances spin lifetimes. However, this hinders the electrical manipulation of electron spins through electric dipole spin resonance (EDSR).\cite{Corna18,Bourdet18} Artificial spin-orbit coupling needs, therefore, to be engineered for that purpose using, e.g., micro-magnets next to the qubits.\cite{Pioro-Ladriere08,Kawakami14,Takeda16}

This fostered the interest in hole qubits that can be efficiently tuned and manipulated electrically owing to the much stronger spin-orbit coupling in the valence band.\cite{Kloeffel13} Electrically driven Rabi oscillations with frequencies up to hundreds of MHz have been reported in hole qubits made of silicon\cite{Maurand16,Crippa18,Crippa19} and germanium.\cite{Watzinger18,Veldhorst19} Two qubit gates have also been demonstrated recently.\cite{Hendrickx20} Modeling shows that hole spins couple to electric fields mostly through ``direct'' spin-orbit interactions in the heavy-hole/light-hole/split-off valence band manifold,\cite{Kloeffel13,Kloeffel18,Venitucci18,Venitucci19} with little contributions from ``remote'' coupling to other bands.\cite{Winkler03}

One of the downside of the strong spin-orbit interactions in the valence band is the increased device-to-device variability and the shorter spin lifetimes resulting from the enhanced coupling to the electrical fluctuations and phonons. It is, therefore, essential to understand the fundamental limits of hole spin-orbit qubits. In this work, we address specifically the coupling between hole spins and phonons. Phonons are indeed intrinsic to the qubits and are much more difficult to mitigate than other scattering mechanisms such as charge noise, which can, in principle, be reduced thanks to a better design of the devices. Recent works have also shown that phonons can dominate the relaxation, especially at high temperature\cite{Loss09,Petit18} and magnetic field. Most studies so far have focused on spin-phonon coupling in electron quantum dots,\cite{Golovach04,Loss05b,Hu14,Tahan14,Petit18} while those on hole quantum dots have made use of simplified strain Hamiltonians or have considered specific geometries.\cite{Woods04,Wu05,Loss05,Heiss07,Loss09,Maier13} {\CORRR We focus here on one-phonon processes, but give equations that account for the full set of deformation potentials and are applicable to any geometry.}

We also discuss two important issues that are also relevant for electron spin qubits. First, we investigate how the spin lifetime depends on the dimensionality of the phonon band structure.\cite{Kargar16} We highlight the differences between bulk-like (3D) and strongly confined 2D and 1D phonons. We show that the transition from the 1D or 2D regime to the 3D regime takes place in ``large'' micrometer size structures because qubits usually couple to low energy phonons that can probe the structure over very long length scales. Second, we explore as a consequence the impact of the encapsulation materials on the spin lifetimes. We show, as a general trend, that the relaxation rate decreases when the qubits are embedded in harder materials.

The paper is organized as follows: in section \ref{sectiontheory}, we present the framework for the calculation of phonon-induced relaxation rates, then discuss the case of hole spin-orbit qubits described by the six bands $\vec{k}\cdot\vec{p}$ model,\cite{Luttinger55,Dresselhaus55,KP09} and the extension to strongly confined 1D and 2D phonons. In section \ref{sectionSOI}, we apply this theory to a hole spin-orbit qubit on silicon-on-insulator (SOI) such as the one measured in Ref. \onlinecite{Crippa18} and modeled in Refs. \onlinecite{Venitucci18} and \onlinecite{Venitucci19}. We identify the dominant contributions to the relaxation, and the optimal operation point for this qubit. We then analyze in section \ref{sectionconfinement} the impact of phonon confinement and of the encapsulation materials on the qubit lifetime.

\section{Theory}
\label{sectiontheory}

In this section, we introduce the general framework for the calculation of the phonon-limited relaxation time, then discuss the application to a prototypical 3D phonon band structure, and to the 6 bands $\vec{k}\cdot\vec{p}$ model for hole spin-orbit qubits. We next extend the formalism to strongly confined 2D and 1D phonon band structures, and highlight, in particular, the different dependences of the relaxation time on the Larmor frequency. We finally discuss the numerical calculation of the acoustic phonon band structure in complex, realistic qubit structures.

\subsection{General framework for 3D phonons}

We consider a qubit based on two eigenstates $\ket{\zero}$ and $\ket{\one}$ of a Hamiltonian $H_0$ (with energies $E_\zero$ and $E_\one$ respectively). {\CORR In a spin-orbit qubit, these two states form a Kramers-degenerate doublet at zero magnetic field (either the ground or an excited one), but the following theory applies to any pair of states.} This qubit interacts with a thermal bath of bulk acoustic phonons with energies $\hbar\omega_{\alpha\vec{q}}$, where $\alpha$ is a branch index and $\vec{q}$ is a 3D wave vector. We assume that the qubit and phonons are coupled by a Hamiltonian $\Delta H[\varepsilon_{ij}(\vec{r})]$ that depends linearly on the local strains $\varepsilon_{ij}(\vec{r})$ ($i,j\in\{x,y,z\}$). These strains can be calculated from the displacement field $\vec{u}(\vec{r})$ of the phonons:\cite{Srivastava90,Boer18} 
\begin{equation}
\varepsilon_{ij}(\vec{r})=\frac{1}{2}\left(\frac{\partial u_i(\vec{r})}{\partial r_j}+\frac{\partial u_j(\vec{r})}{\partial r_i}\right)\,.
\label{eqstrain}
\end{equation}
The displacement field operator in branch $\alpha$ further reads:
\begin{equation}
\vec{u}_{\alpha\vec{q}}(\vec{r},t)=A_{\alpha\vec{q}}(\vec{r},t)\hat{\vec{c}}_{\alpha\vec{q}}\,,
\label{equphonons}
\end{equation}
where $\hat{\vec{c}}_{\alpha\vec{q}}$ is the unit phonon polarization vector and:
\begin{equation}
A_{\alpha\vec{q}}(\vec{r},t)=\sqrt{\frac{\hbar}{2\rho\Omega\omega_{\alpha\vec{q}}}}e^{i\vec{q}\cdot\vec{r}}(a_{\alpha\vec{q}}e^{-i\omega_{\alpha\vec{q}}t}+a^\dagger_{\alpha,\vec{-q}}e^{i\omega_{\alpha\vec{q}}t})\,,
\end{equation}
with $\rho$ the density of the host material, $\Omega$ the volume of the system, and $a^\dagger_{\alpha\vec{q}}$ the phonon creation operators. The strain tensor that derives from Eq. (\ref{equphonons}) is therefore:
\begin{equation}
\varepsilon_{\alpha\vec{q}}(\vec{r},t)=iqA_{\alpha\vec{q}}(\vec{r},t)\epsilon_{\alpha\vec{q}}\,,
\end{equation}
where:
\begin{equation}
\epsilon_{\alpha\vec{q}}=\frac{1}{2}
\begin{bmatrix}
  2\hat{c}_x \hat{q}_x & \hat{c}_x \hat{q}_y + \hat{c}_y \hat{q}_x & \hat{c}_x \hat{q}_z + \hat{c}_z \hat{q}_x \\
  \hat{c}_y \hat{q}_x + \hat{c}_x \hat{q}_y & 2\hat{c}_y \hat{q}_y & \hat{c}_y \hat{q}_z + \hat{c}_z \hat{q}_y \\
  \hat{c}_z \hat{q}_x + \hat{c}_x \hat{q}_z & \hat{c}_z \hat{q}_y + \hat{c}_y \hat{q}_z & 2\hat{c}_z \hat{q}_z
\end{bmatrix}\,,
\end{equation}
with $\hat{\vec{c}}\equiv\hat{\vec{c}}_{\alpha\vec{q}}$ and $\hat{\vec{q}}$ the unit vector along $\vec{q}$.

\begin{widetext}
Assuming $E_{\one}>E_{\zero}$, the rate of transitions $\Gamma_{\zero\one}^{\rm 3D}$ from state $\ket{\zero}$ to state $\ket{\one}$ due to phonon absorption is given  
by Fermi's golden rule:\cite{Woods04,Wu05,Tahan14}
\begin{equation}
\Gamma_{\zero\one}^{\rm 3D}=\frac{2\pi}{\hbar}\sum_{\alpha}\int d^3\vec{q}\, \rho_{\vec{q}}
\frac{\hbar q^2 N_{\alpha\vec{q}}}{2 \rho \Omega \omega_{\alpha\vec{q}}} 
\Big| \langle \zero | e^{i \vec{q}\cdot\vec{r}} \Delta H(\epsilon_{\alpha\vec{q}}) | \one \rangle \Big|^2 \delta(E_\one - E_\zero - \hbar\omega_{\alpha\vec{q}})\,,
\label{eqgoldenrule01}
\end{equation}
where $\rho_{\vec{q}}=\Omega/(2\pi)^3$ is the density of states in reciprocal space, and $N_{\alpha\vec{q}}=1/(e^{\beta\hbar\omega_{\alpha\vec{q}}}-1)$ is the thermal population of phonons with energy $\hbar\omega_{\alpha\vec{q}}$ (with $\beta=1/(k_BT)$ and $T$ the temperature). This equation accounts exclusively for one-phonon processes and is, therefore, valid only at low temperature.\cite{Loss09,Petit18} The rate of transitions $\Gamma_{\one\zero}^{\rm 3D}$ from state $\ket{\one}$ to state $\ket{\zero}$ due to spontaneous and stimulated phonon emission is given by the same expression with $N_{\alpha\vec{q}}$ replaced by $N_{\alpha\vec{q}}+1$. Therefore, the total relaxation rate $\Gamma_{\rm ph}^{\rm 3D}=T_1^{-1}=\Gamma^{\rm 3D}_{\zero\one}+\Gamma^{\rm 3D}_{\one\zero}$ reads:
\begin{equation}
\Gamma_{\rm ph}^{\rm 3D}=\frac{1}{8\pi^2\rho\hbar\omega}\coth\left(\frac{\hbar\omega}{2k_BT}\right)\sum_{\alpha}\int d^3\vec{q}\, 
q^2 \Big| \langle \zero | e^{i \vec{q}\cdot\vec{r}} \Delta H(\epsilon_{\alpha\vec{q}}) | \one \rangle \Big|^2 \delta(\omega - \omega_{\alpha\vec{q}})\,,
\label{eqgoldenrule}
\end{equation}
Only phonons matching the Larmor frequency $\omega=(E_\one-E_\zero)/\hbar$ couple to the qubit as highlighted by the delta function. This equation can be transformed as:
\begin{equation}
\Gamma_{\rm ph}^{\rm 3D}=\frac{1}{8\pi^2\rho\hbar\omega}\coth\left(\frac{\hbar\omega}{2k_BT}\right)\sum_{\alpha}\int_{S_\alpha(\omega)} d^2\vec{q}\,
\frac{q^2}{|\vec{v}_{\alpha\vec{q}}|} 
\Big| \langle \zero | e^{i \vec{q}\cdot\vec{r}} \Delta H(\epsilon_{\alpha\vec{q}}) | \one \rangle \Big|^2\,,
\label{eqgoldenrule2}
\end{equation}
where $S_\alpha(\omega)$ is the surface $\omega_{\alpha\vec{q}}=\omega$ and $\vec{v}_{\alpha\vec{q}}=\vec{\nabla}_{\vec{q}}\omega_{\alpha\vec{q}}$ is the group velocity of the phonons. The integration can be finalized once a model has been chosen for the phonon and electronic band structures of the qubit. 
\end{widetext}

\subsection{Model 3D phonon band structure}

In the following, we consider a prototypical, isotropic 3D phonon band structure with one longitudinal acoustic (LA) branch $\omega_{l\vec{q}}=v_lq$, and two degenerate transverse acoustic (TA1 and TA2) branches $\omega_{t_1\vec{q}}=\omega_{t_2\vec{q}}=v_tq$, where $v_l$ and $v_{t_1}=v_{t_2}=v_t$ are the longitudinal and transverse sound velocities. We specify the orientation of the phonon wave vector by the azimuthal and polar angles $\theta$ and $\varphi$ so that $\hat{\vec{q}}=(\sin\theta\cos\varphi,\sin\theta\sin\varphi,\cos\theta)$. The polarization vector of the LA branch $\hat{\vec{c}}_l=\hat{\vec{q}}$ can, therefore, be characterized by the angles ($\theta_l=\theta$, $\varphi_l=\varphi$), while the polarization vectors $\hat{\vec{c}}_{t_1}$ and $\hat{\vec{c}}_{t_2}$ of the TA branches can be characterized by the angles ($\theta_{t_1}=\theta+\pi/2$, $\varphi_{t_1}=\varphi$) and ($\theta_{t_2}=\pi/2$, $\varphi_{t_2}=\varphi+\pi/2$), respectively.
Then,
\begin{widetext}
\begin{equation}
\Gamma_{\rm ph}^{\rm 3D}=\frac{\omega^3}{8\pi^2\hbar\rho} \coth\left(\frac{\hbar\omega}{2k_BT}\right) \sum_{\alpha\in{l,t_1,t_2}} \frac{1}{v_\alpha^5} \int_{0}^{\pi}d\theta\,\sin\theta\int_{0}^{2\pi}d\varphi\,\Big|\langle \zero | e^{iq_\alpha\hat{\vec{q}}(\theta,\varphi)\cdot\vec{r}} \Delta H[\epsilon_{\alpha}(\theta,\varphi)]| \one \rangle \Big|^2,
\label{eqgamma3D}
\end{equation}
where $v_{\alpha}q_{\alpha}=\omega$, and:
\begin{subequations}
\begin{equation}
\epsilon_l(\theta,\varphi) = \frac{1}{2}
\begin{bmatrix}
    2\sin^2\theta \cos^2\varphi & \sin^2\theta \sin 2\varphi & \sin 2\theta \cos\varphi \\
    \sin^2\theta \sin 2\varphi & 2\sin^2\theta \sin^2\varphi & \sin 2\theta \sin\varphi \\
    \sin 2\theta \cos\varphi & \sin 2\theta \sin\varphi & 2\cos^2\theta 
\end{bmatrix}
\end{equation}
\begin{equation}
\epsilon_{t_1}(\theta,\varphi) = \frac{1}{2}
\begin{bmatrix}
    \sin 2\theta \cos^2\varphi & \frac{1}{2} \sin 2\theta \sin 2\varphi & \cos 2\theta \cos\varphi \\
    \frac{1}{2} \sin 2\theta \sin 2\varphi & \sin 2\theta \sin^2\varphi & \cos 2\theta \sin\varphi \\
    \cos 2\theta \cos\varphi & \cos 2\theta \sin\varphi & -\sin 2\theta 
\end{bmatrix}
\label{eqstrainTA1}
\end{equation}
\begin{equation}
\epsilon_{t_2}(\theta,\varphi) = \frac{1}{2} 
\begin{bmatrix}
   -\sin\theta \sin 2\varphi & \sin\theta \cos 2\varphi & -\cos\theta \sin\varphi  \\
    \sin\theta \cos 2\varphi & \sin\theta \sin 2\varphi &  \cos\theta \cos\varphi \\
   -\cos\theta \sin\varphi & \cos\theta \cos\varphi &  0
\end{bmatrix}\,.
\label{eqstrainTA2}
\end{equation}
\end{subequations}
\end{widetext}
In the single band effective mass approximation for silicon, we recover the expressions of Ref. \onlinecite{Tahan14} for the phonon-induced relaxation rate. We further discuss the case of hole qubits within the six bands $\vec{k}\cdot\vec{p}$ model in the next subsection.

\subsection{Application to the six bands $\vec{k}\cdot\vec{p}$ model}

In the six bands $\vec{k}\cdot\vec{p}$ model, the hole wave functions are expanded as:\cite{Luttinger55,Dresselhaus55,KP09}
\begin{align}
\psi(\vec{r})&=F_{X\uparrow}(\vec{r})u_{X\uparrow}(\vec{r})+F_{X\downarrow}(\vec{r})u_{X\downarrow}(\vec{r}) \nonumber \\
&+F_{Y\uparrow}(\vec{r})u_{Y\uparrow}(\vec{r})+F_{Y\downarrow}(\vec{r})u_{Y\downarrow}(\vec{r}) \nonumber \\
&+F_{Z\uparrow}(\vec{r})u_{Z\uparrow}(\vec{r})+F_{Z\downarrow}(\vec{r})u_{Z\downarrow}(\vec{r})\,,
\label{eqsixbands}
\end{align}
where the $F_{i\sigma}$'s are envelope functions and the $u_{i\sigma}$'s are the Bloch functions at $\Gamma$. They are bonding combinations of atomic $p_X$, $p_Y$ or $p_Z$ orbitals with spin $\sigma=\,\uparrow$ or $\downarrow$ along $\vec{Z}$ (the $\vec{X}\parallel[100]$, $\vec{Y}\parallel[010]$, $\vec{Z}\parallel[001]$ axes being the cubic axes).

Alternatively the $u$'s (hence the $F$'s) can be mapped by a unitary transform onto the eigenstates $\ket{J, m_J}$ of $\vec{J}^2$ and $J_{\vec{u}}$, where $\vec{J}=\vec{L}+\vec{S}$ is the total angular momentum of the Bloch functions and $\vec{u}$ is an arbitrary quantization axis. Actually, any six bands ($J=3/2$ \& $J=1/2$ multiplets) or four bands ($J=3/2$ multiplet) flavor of the $\vec{k}\cdot\vec{p}$ model for the valence band can be put in the form of Eq. (\ref{eqsixbands}).\cite{Luttinger55,Dresselhaus55,KP09} In the $\{\ket{X\uparrow}, \ket{Y\uparrow}, \ket{Z\uparrow}, \ket{X\downarrow}, \ket{Y\downarrow}, \ket{Z\downarrow}\}$ basis set, the Hamiltonian $\Delta H(\varepsilon)$ is:\cite{Bir59,Bahder90,KP09}
\begin{equation}
\Delta H(\varepsilon)=
\begin{bmatrix}
 \Delta H_s(\varepsilon) & 0_{3\times 3} \\
 0_{3\times 3} & \Delta  H_s(\varepsilon)
\end{bmatrix}\,,
\label{eqdeltaH}
\end{equation}
where the $3\times 3$ up or down spin sub-block $\Delta H_s(\varepsilon)$ reads:
\begin{widetext}
\begin{equation}
\Delta H_s(\varepsilon)=
\begin{bmatrix}
 l \varepsilon_{XX} + m \varepsilon_{YY} + m \varepsilon_{ZZ} & n \varepsilon_{XY} & n \varepsilon_{XZ} \\
 n \varepsilon_{XY} & m \varepsilon_{XX} + l \varepsilon_{YY} + m \varepsilon_{ZZ} & n \varepsilon_{YZ} \\
 n \varepsilon_{XZ} & n \varepsilon_{YZ} & m \varepsilon_{XX} + m \varepsilon_{YY} + l \varepsilon_{ZZ} 
\end{bmatrix}
\label{eqdeltaHs}
\end{equation}
\end{widetext}
with:
\begin{eqnarray}
l=a+2b\,; \hspace{0.2cm}
m=a-b\,; \hspace{0.2cm}
n=d\sqrt{3}\,.
\end{eqnarray}
$a\equiv a_v$ is the hydrostatic deformation potential, $b$ the uniaxial deformation potential, and $d$ the shear deformation potential of the valence band.\cite{noteDelta} The above model accounts for spin relaxation due to ``direct'' spin-orbit interactions within the valence band manifold;\cite{Kloeffel11,Kloeffel13,Kloeffel18} Even if $\Delta H$ is diagonal in spin, it can drive relaxation between textured spin states such as the mixed heavy- and light-hole states encountered in spin-orbit qubits.

We next write:
\begin{subequations}
\begin{align}
\langle\vec{r}|\zero\rangle&=\sum_{i\in\{X,Y,Z\}}\sum_{\sigma\in\{\uparrow,\downarrow\}} a_{i\sigma}(\vec{r})u_{i\sigma}(\vec{r}) \\
\langle\vec{r}|\one\rangle&=\sum_{i\in\{X,Y,Z\}}\sum_{\sigma\in\{\uparrow,\downarrow\}} b_{i\sigma}(\vec{r})u_{i\sigma}(\vec{r})\,,
\end{align}
\end{subequations}
{\CORR and assume that the extension of the envelope functions $a_{i\sigma}(\vec{r})$ and $b_{i\sigma}(\vec{r})$ is much smaller than the typical wave length of the phonons at the Larmor frequency. We can then complete the integration of Eq. (\ref{eqgamma3D}) in the dipole approximation for the phase factor $e^{i\vec{q}\cdot\vec{r}}$ (see Appendix \ref{appendixdipole}), and get:}
\begin{align}
\Gamma_{\rm ph}^{\rm 3D}&=\frac{\omega^3}{8\pi\hbar\rho}\coth\left(\frac{\hbar\omega}{2k_BT}\right) \nonumber \\
&\times\sum_{\alpha=l,t}\left(\frac{\omega^2}{v_\alpha^7}\sum_{n=1}^9 A_n\Lambda^{\rm A}_{n\alpha} + \frac{1}{v_\alpha^5}\sum_{n=1}^2 B_n\Lambda^{\rm B}_{n\alpha}\right)\,,
\label{eq3Drelax}
\end{align}
where the $\Lambda$'s depend on material parameters and the $A_n$'s and $B_n$'s can be expressed as a function of the following moments of the hole envelopes:
\begin{subequations}
\begin{align}
S_{ij}&=\int d^3\vec{r}\, \left[a^*_{i\uparrow}(\vec{r}) b_{j\uparrow}(\vec{r}) + a^*_{i\downarrow}(\vec{r}) b_{j\downarrow}(\vec{r})\right] \\
R^k_{ij}&=\int d^3\vec{r}\, \left[a^*_{i\uparrow}(\vec{r}) b_{j\uparrow}(\vec{r}) + a^*_{i\downarrow}(\vec{r}) b_{j\downarrow}(\vec{r})\right]r_k \\
T^{kk'}_{ij}&=\int d^3\vec{r}\, \left[a^*_{i\uparrow}(\vec{r}) b_{j\uparrow}(\vec{r}) + a^*_{i\downarrow}(\vec{r}) b_{j\downarrow}(\vec{r})\right]r_k r_{k'}
\end{align}
\label{eqmoments}
\end{subequations}
and:
\begin{equation}
O^{mn}_{ijkl}=R^m_{ij}R^{n*}_{kl}-\frac{1}{2}\left(T^{mn}_{ij}S_{kl}^* + T^{mn*}_{kl}S_{ij}\right)\,.
\label{eqOmijkl}
\end{equation}
Namely,
\begin{subequations}
\begin{align}
A_1&=\sum_{i}O^{ii}_{iiii} \\
A_2&=\sum_{i \neq j} O^{ii}_{iijj}+O^{ii}_{jjii} \\
A_3&=\sum_{i \neq j} O^{ii}_{jjjj} \\
A_4&=\sum_{i \neq j \neq k} O^{ii}_{jjkk} \\
A_5&=\sum_{i \neq j} O^{ii}_{ijij}+O^{ii}_{ijji}+O^{ii}_{jiij}+O^{ii}_{jiji} \\
A_6&=\sum_{i \neq j \neq k} O^{ii}_{jkjk}+O^{ii}_{jkkj} 
\end{align}
\begin{align}
A_7&=\sum_{i \neq j} O^{ij}_{iiij}+O^{ij}_{iiji}+O^{ij}_{ijii}+O^{ij}_{jiii}+\rm{c.c.} \\
A_8&=\sum_{i \neq j \neq k} O^{ij}_{ijkk}+O^{ij}_{jikk}+\rm{c.c.} \\
A_9&=\sum_{i \neq j \neq k} O^{ij}_{ikjk}+O^{ij}_{kijk}+O^{ij}_{ikkj}+O^{ij}_{kikj} \nonumber \\ 
& \ \ \ \ \ \ \ \  +O^{ij}_{jkik}+O^{ij}_{kjik}+O^{ij}_{jkki}+O^{ij}_{kjki}\,,
\end{align}
\end{subequations}
where c.c. stands for complex conjugate, and:
\begin{subequations}
\begin{align}
B_1&=\sum_{i} S_{ii}S^*_{ii} \\
B_2&=\sum_{i\neq j} S_{ij}(S^*_{ij}+S^*_{ji})\,.
\label{eqBs}
\end{align}
\end{subequations}
The sums over $i,j,k$ run over $\{X,Y,Z\}$. The $\Lambda^{\rm A}_{nl}$ parameters for the longitudinal phonons are:
\begin{subequations}
\begin{align}
\Lambda^{\rm A}_{1l}&=\frac{140 a^2 + 224 ab + 176 b^2}{105} \\
\Lambda^{\rm A}_{2l}&=\frac{140 a^2 + 56 ab - 88 b^2}{105} \\
\Lambda^{\rm A}_{3l}&=\frac{140 a^2 - 112 ab + 80 b^2}{105} \\
\Lambda^{\rm A}_{4l}&=\frac{140 a^2 - 112 ab + 8 b^2}{105} \\
\Lambda^{\rm A}_{5l}&=\frac{12 d^2}{35} \\
\Lambda^{\rm A}_{6l}&=\frac{4 d^2}{35} \\
\Lambda^{\rm A}_{7l}&=\frac{28 ad + 8 bd}{35\sqrt{3}} \\ 
\Lambda^{\rm A}_{8l}&=\frac{28 ad - 16 bd}{35\sqrt{3}} \\
\Lambda^{\rm A}_{9l}&=\frac{4 d^2}{35} 
\end{align}
\end{subequations}
while the $\Lambda^{\rm A}_{nt}$ parameters for transverse phonons are:
\begin{subequations}
\begin{align}
\Lambda^{\rm A}_{1t}&=\frac{72 b^2}{35} \\
\Lambda^{\rm A}_{2t}&=-\frac{36 b^2}{35} \\
\Lambda^{\rm A}_{3t}&=\frac{48 b^2}{35} \\
\Lambda^{\rm A}_{4t}&=-\frac{12 b^2}{35} \\ 
\Lambda^{\rm A}_{5t}&=\frac{16 d^2}{35} \\
\Lambda^{\rm A}_{6t}&=\frac{2 d^2}{7}
\end{align}
\begin{align}
\Lambda^{\rm A}_{7t}&=\frac{2\sqrt{3} bd }{35} \\
\Lambda^{\rm A}_{8t}&=-\frac{4\sqrt{3} bd }{35} \\
\Lambda^{\rm A}_{9t}&=\frac{3 d^2}{35}\,. 
\end{align}
\end{subequations}
Finally, the $\Lambda^{\rm B}_{nl}$ and $\Lambda^{\rm B}_{nt}$ parameters are:
\begin{subequations}
\begin{align}
\Lambda^{\rm B}_{1l}&=\frac{24 b^2}{5} \\
\Lambda^{\rm B}_{2l}&=\frac{4 d^2}{5}\,,
\end{align}
\end{subequations}
\begin{subequations}
\begin{align}
\Lambda^{\rm B}_{1t}&=\frac{36 b^2}{5} \\
\Lambda^{\rm B}_{2t}&=\frac{6 d^2}{5}\,.
\end{align}
\end{subequations}
The TA1 and TA2 branches have been summed up in the transverse phonon parameters $\Lambda^{\rm A}_{nt}$ and $\Lambda^{\rm B}_{nt}$. The two $B_n$ terms are ``band mixing'' terms whose prefactor scales with the phonon strains ($\propto q^2/\omega\equiv\omega$) and density of states ($\propto\omega^2$) at the Larmor frequency. They result from the zero-th order expansion $e^{i\vec{q}\cdot\vec{r}}\sim 1$ in Eq. (\ref{eqgoldenrule2}) (homogeneous component of the strain), and are ruled by the overlap between the different envelope functions. The 9 $A_n$ terms are dipole-like terms that follow from the first-order expansion of $e^{i\vec{q}\cdot\vec{r}}$ in Eq. (\ref{eqgoldenrule2}). They show, therefore, an additional $\propto q^2\equiv\omega^2$ dependence. Although the $A_n$ terms also mix bands, we point out that the $B_n$ terms do not exist in one-band models where the envelopes of different states are, by design, orthogonal. Note that the $A_n$ and $B_n$ terms are also dependent on $\omega$ (see next section). We will sort the different contributions in section \ref{sectionterms}.

\subsection{Model 2D \& 1D phonon band structures}
\label{section1D2D}

The 3D model discussed up to now is suitable for ``large'' enough structures where the phonons (yet not necessarily the electrons or holes) are weakly confined (bulk-like dispersion). In confined structures such as nanowires or thin films, the vibrational modes may, however, be more adequately described by a 1D or 2D phonon band structure.\cite{Bannov95,Nishiguchi97,Thonhauser04,Kargar16} The integration over $\varphi$ and/or $\theta$ in Eq. (\ref{eqgamma3D}) is then replaced by a sum over phonon sub-bands. In general, many sub-bands can contribute to the relaxation rate. Yet the splitting between phonon sub-bands increases with lateral confinement so that only the 1D or 2D acoustic branches will ultimately couple to the qubit once the phonons get confined enough. The conditions in which this strongly confined regime is achieved will be explored in section \ref{sectionbulklimit}.

We can derive expressions similar to Eq. (\ref{eq3Drelax}) in the strongly confined regime by making simple assumptions for the acoustic phonon wave functions. Namely, we assume that the displacements are homogeneous in the thickness of the film (2D) or in the cross section of the nanowire (1D). This corresponds to a choice of periodic Born-von-Karman instead of free-standing boundary conditions at the surface of the film or wire, or equivalently to a sampling of the 3D phonon band structure at $\vec{q}_\perp=\vec{0}$, where $\vec{q}_\perp$ is the component of the wave vector perpendicular to the film or wire. The relevance and validity of this approximation will be discussed in section \ref{sectionbulklimit}. We then reach the following expressions for the relaxation rate in thin films (see Supplementary Material):
\begin{widetext}
\begin{equation}
\Gamma_{\rm ph}^{\rm 2D} = \frac{\omega ^2}{4\hbar\rho L} \coth\left(\frac{\hbar\omega}{2k_BT}\right) \sum_{\alpha=l,t} \left(\frac{\omega^2}{v_{\alpha}^6} \sum_{n} A_n \Lambda^{\rm A}_{n\alpha} + \frac{1}{v_{\alpha}^4} \sum_{n} B_n \Lambda^{\rm B}_{n\alpha} \right)\,,
\label{eq2Drelax}
\end{equation}
where $L$ is the thickness of the film, and in nanowires:
\begin{equation}
\Gamma_{\rm ph}^{\rm 1D} = \frac{\omega}{2\hbar\rho S} \coth\left(\frac{\hbar\omega}{2k_BT}\right) \sum_{\alpha=l,t} \left(\frac{\omega^2}{v_{\alpha}^5} \sum_{n} A_n \Lambda^{\rm A}_{n\alpha} + \frac{1}{v_{\alpha}^3} \sum_{n} B_n \Lambda^{\rm B}_{n\alpha} \right)\,,
\label{eq1Drelax}
\end{equation}
where $S$ is the cross sectional area of the wire.
\end{widetext}
The $A_n$'s, $B_n$'s and $\Lambda$'s depend on the dimensionality and on the crystallographic orientation of the nanostructure. They are given for $(100)$, $(110)$, and $(111)$ films, and for $[100]$, $[110]$, and $[111]$ oriented wires in the Supplementary Material.

With respect to bulk phonons [Eq. (\ref{eq3Drelax})], the relaxation rate is inversely proportional to the characteristic size of the system, and the dependence on $\omega$ is reduced by one power each time the phonons get confined in an additional direction. This results from the scaling of the phonon density of states with dimensionality. We emphasize, however, that the net dependence on the relaxation rates on the Larmor frequency $\omega$ is typically two orders of magnitudes stronger than suggested by Eqs. (\ref{eq3Drelax}), (\ref{eq2Drelax}) and (\ref{eq1Drelax}), because the $A_n$'s and $B_n$'s also depend on $\omega$. Indeed, they vanish at zero magnetic field if $\ket{\zero}$ and $\ket{\one}$ form a Kramers degenerate pair linked by time-reversal symmetry ({\CORR as is the case in a spin-orbit qubit}), and increase as $\omega^2\propto B^2$ once a magnetic field $\vec{B}$ breaks time-reversal symmetry. 

Also, Eqs. (\ref{eq3Drelax}), (\ref{eq2Drelax}) and (\ref{eq1Drelax}) can actually be used to calculate the transition rates between any pair of states ({\CORR provided the dipole approximation applies at the transition energy}), in order to set up master equations for spin and charge relaxation in the system for example.\cite{Loss05,Kornich18} The coth term shall be replaced with $N(\omega)=1/(e^{\beta\hbar\omega}-1)$ for absorption rates and by $N(\omega)+1$ for emission rates.

\subsection{Numerical phonon band structures}
\label{sectionnumerical}

Real {\CORR qubit} devices are often made of nanostructured stacks of materials with complex phonon band structures. In order to validate the above models and address, e.g., the impact of encapsulation materials on the relaxation rates, we have developed a numerical approach to the acoustic phonon band structure based on continuum elasticity theory.\cite{Srivastava90,Nishiguchi97,Boer18} {\CORRR The latter is expected to hold at small strains in semiconductor heterostructures with characteristic sizes around 10 nm;\cite{Bernard94,Pryor98} Non-local and surface effects have been shown to change the bending rigidity of 10 nm $\times$ 10 nm Si and SiO$_2$ beams by less than 10\%.\cite{Maranganti07}} We consider, in particular, 1D structural models with arbitrary cross-section and materials. This cross-section is meshed and the dynamical matrix is computed from a finite-element discretization of the continuum elasticity theory.

The relaxation rate is still calculated from Fermi's golden rule: 
\begin{align}
\Gamma_{\rm ph}&=\frac{1}{2\hbar\omega}\coth\left(\frac{\hbar\omega}{2k_BT}\right) \nonumber \\
&\times\sum_{\alpha}\sum_{q\in\{q_\alpha\}} 
\frac{1}{|v_{\alpha q}|} 
\Big| \langle \zero | \Delta H[\tilde{\varepsilon}_{\alpha q}(\vec{r})] | \one \rangle \Big|^2\,,
\label{eqGammanum}
\end{align}
where $\alpha$ run over all 1D sub-bands, $\{q_\alpha\}$ are the solutions of $\omega_{\alpha q_{\alpha}}=\omega$ (if any), and:
\begin{equation}
[\tilde{\varepsilon}_{\alpha q}(\vec{r})]_{ij}=\frac{1}{2}\left(\frac{\partial\tilde{u}_i(\vec{r})}{\partial r_j}+\frac{\partial\tilde{u}_j(\vec{r})}{\partial r_i}\right)
\end{equation}
with: 
\begin{equation}
\tilde{\vec{u}}(\vec{r})=\frac{1}{\sqrt{\rho(y,z)}}\vec{c}(y,z)e^{iqx}
\end{equation} 
and $\vec{c}\equiv\vec{c}_{\alpha q}$ the eigenvector of the dynamical matrix (normalized so that $\int dydz\,|\vec{c}(y,z)|^2=1$). The 1D wave vector is aligned along the $\vec{x}$ axis. Note that we do not make the dipole approximation on the phase factor here. The convergence of Eq. (\ref{eqGammanum}) is discussed in Appendix \ref{appendixConvergence}.

\section{Application to hole-spin qubits on silicon-on-insulator}
\label{sectionSOI}

\begin{figure}
\includegraphics[width=1.0\columnwidth]{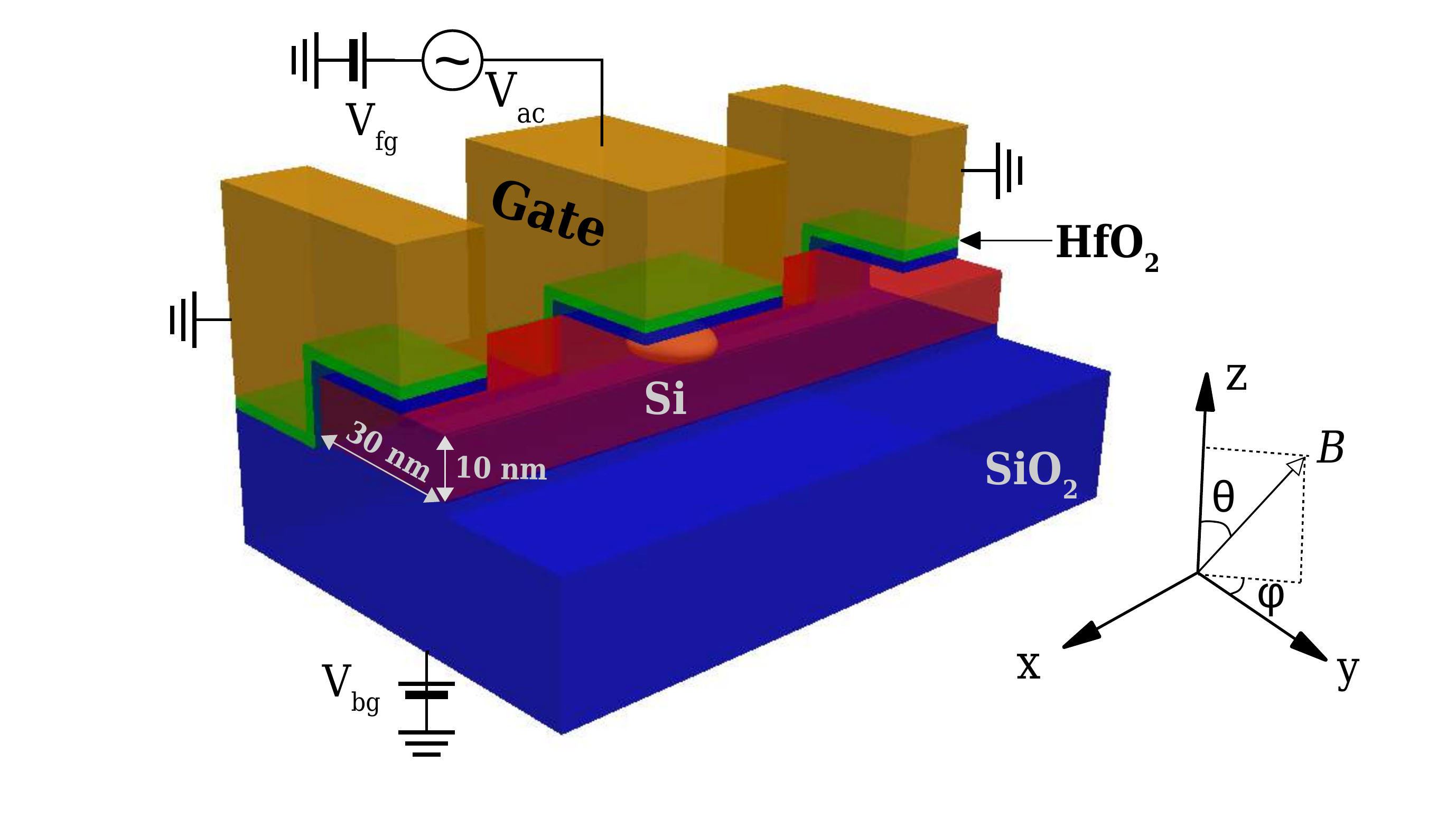}
\caption{Schematics of the hole qubit device on silicon-on-insulator: a silicon nanowire channel (red) with width $W=30$ nm and height $H=10$ nm lies on top of the buried oxide (blue). Top gates (orange) with length $L_g=30$ nm and spacing $S_g=30$ nm partly cover the silicon nanowire (over $20$ nm). The gate stack is made of 2 nm of SiO$_2$ and 2 nm of HfO$_2$ (green). The central front gate defines the hole quantum dot. An iso-density surface of the confined hole ground-state wave function is depicted in yellow. The device axes are defined on the figure.}
\label{figdevice}
\end{figure}

In this section, we apply the models introduced in section \ref{sectiontheory} to a hole spin-orbit qubit on silicon-on-insulator. The device, represented in Fig. \ref{figdevice}, is the same as the one modeled in Ref. \onlinecite{Venitucci18}, and is similar to the one measured in Ref. \onlinecite{Crippa18}. It is made of a $[110]$-oriented silicon nanowire channel with width $W=30$ nm [$(1\bar{1}0)$ facets] and height $H=10$ nm [$(001)$ facets] lying on top of a 25 nm thick buried oxide and silicon substrate. A 30 nm long central gate partly overlapping the nanowire controls an electrostatically defined quantum dot. The two other gates on the left and right mimic neighboring qubits. The gate stack is made of 2 nm of SiO$_2$ and 2 nm of HfO$_2$. The whole device is embedded in Si$_3$N$_4$. The qubit can be further controlled using the substrate as a back gate, which allows to tune separately the chemical potential and the electric field in the dot. We bias the central front gate at $V_{\rm fg}=-0.1$ V and ground the other front gates in order to confine holes. Rabi oscillations can be electrically driven by a radio-frequency signal on the central front gate.

The potential landscape in the device is computed with a finite volumes Poisson solver and the qubit states (chosen as the topmost valence band states) with a finite differences discretization of the six bands $\vec{k}\cdot\vec{p}$ model.\cite{Venitucci18} {\CORRR For the model phonon band structures, the group velocities are $v_l=9000$ m/s and $v_t=5400$ m/s. The numerical phonon band structures are computed with the elastic constants listed in Table \ref{tablecij}. The deformation potentials of the valence band of silicon are $a=2.38$ eV, $b=-2.1$ eV, and $d=-4.85$ eV.\cite{Laude71,Li06} The amplitude of the magnetic field is adjusted so that the Larmor frequency of the qubit sticks to $\omega/(2\pi)=10$ GHz. {\CORR At that frequency, the wave length of bulk acoustic phonons is greater than $\lambda_t=2\pi v_t/\omega=540$ nm, hence is much longer than the dot size ($\sim 30$ nm) and within the range of validity of the dipole approximation.} {\CORRRR The temperature is set to $T=100$ mK. The results are, however, weakly dependent on temperature $T\le 100$ mK, as $1<\coth(\hbar\omega/2k_BT)<1.017$ in this range (base cryostat temperature was $T\simeq 15$ mK in Refs. \onlinecite{Maurand16} and \onlinecite{Crippa18}). Also, the splitting between the ground qubit states and the first excited orbital states is always larger than 2 meV in this device (Fig. 8 of Ref. \onlinecite{Venitucci18}). From the scaling laws given in Ref. \onlinecite{Loss09}, we do not expect significant two-phonon corrections at such low temperatures, although we did not compute them explicitly at this stage.}

As discussed in Ref. \onlinecite{Venitucci18}, the Larmor and Rabi frequencies of this device are strongly dependent on the orientation of the magnetic field and on the back gate voltage $V_{\rm bg}$. This is a fingerprint of the action of the spin-orbit coupling on the holes, which (in the absence of strains) have a dominant heavy-hole character along the strong confinement axis $z=[001]$.\cite{Kloeffel11,Kloeffel13,Kloeffel18} In particular, the Rabi frequency exhibits a dip at back gate voltage $V_{\rm bg}\simeq-0.15$ V where the hole wave functions show an approximate inversion center that hampers the action of spin-orbit coupling (the hole spins decouple from the gate electric field). This will be further investigated in section \ref{sectionoptimal}.

We first discuss which terms do play a role in Eq. (\ref{eq3Drelax}), then the optimal bias point and magnetic field orientation for this qubit. 

\begin{table}
\begin{tabular}{l d d d d}
\hline \hline
Material & \multicolumn{1}{c}{$\rho$} & \multicolumn{1}{c}{$c_{11}$} & \multicolumn{1}{c}{$c_{12}$} & \multicolumn{1}{c}{$c_{44}$} \\
\hline
Si (aniso)\cite{McSkimin64} & 2.329 &  166.0 &  64.0 &  79.6 \\
Si (iso) & 2.329 &  188.6 &  52.8 &  67.9 \\
SiO$_2$\cite{Pabst13} & 2.200 &   77.5 &  15.7 &  30.9 \\
Si$_3$N$_4$\cite{Kramer03,Ziebart97} & 2.500 &  193.0 &  65.0 &  64.0 \\
Diamond\cite{Grimsditch75} & 3.500 & 1076.0 & 125.0 & 577.0 \\
\hline
\hline
\end{tabular}
\caption{Density $\rho$ (g/cm$^3$) and elastic constants $c_{11}$, $c_{12}$ and $c_{44}$ (GPa) of the different materials considered in this work.\cite{McSkimin64,Pabst13,Kramer03,Ziebart97,Grimsditch75} The anisotropic (aniso) elastic constants of silicon are those measured in this material.\cite{McSkimin64} They give rise to anisotropic phonon bands with two non-degenerate transverse acoustic branches in bulk. The isotropic (iso) elastic constants have been modified in order to get isotropic phonon bands with degenerate transverse acoustic branches consistent with the model 3D phonon band structure ($v_l=9000$ m/s and $v_t=5400$ m/s). The elastic models for amorphous SiO$_2$ and Si$_3$N$_4$ are isotropic ($c_{11}-c_{12}=2c_{44}$).}
\label{tablecij}
\end{table}

\subsection{Which terms do play a role ?}
\label{sectionterms}

\begin{figure*}
\includegraphics[width=1.8\columnwidth]{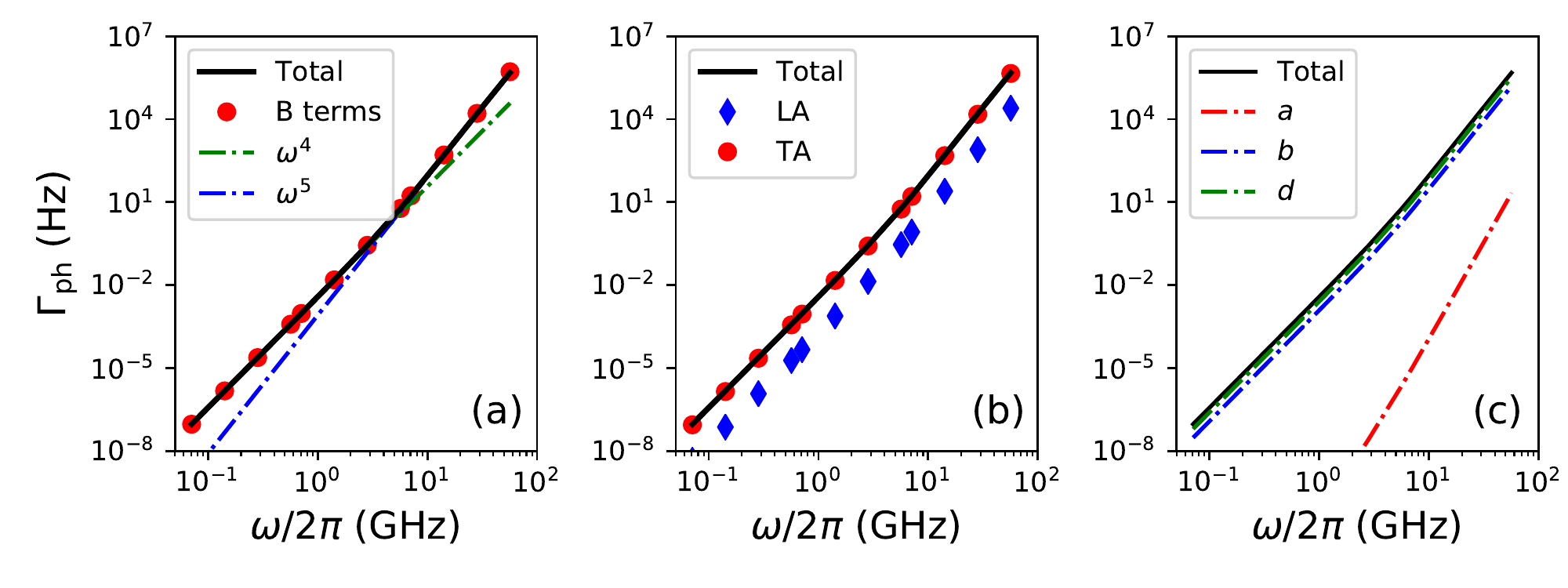}
\includegraphics[width=1.8\columnwidth]{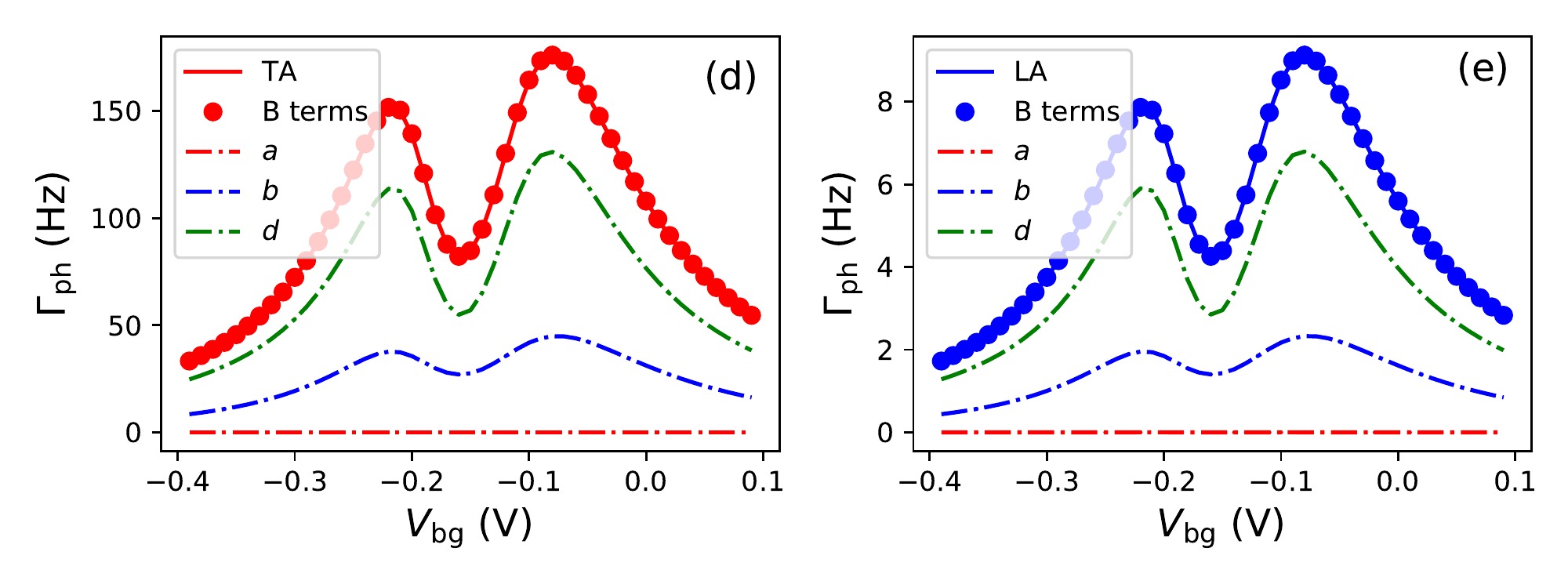}
\caption{(a-c) Contribution of the different terms of Eq.~(\ref{eq3Drelax}) to the relaxation rate as a function of the Larmor frequency $\omega$ ($V_{\rm bg}=-0.15$ V). The relaxation is dominated by (a) the homogeneous band mixing $B_n$ terms, (b) the transverse acoustic phonons, and (c) the shear deformation potential $d$. The above conclusions are valid in a wide range of back gate voltages $V_{\rm bg}$, as shown in panels (d) and (e). The orientation of the magnetic field is $\theta=45^\circ$, $\varphi=0$ in all panels and its amplitude is adjusted so that the Larmor frequency is $\omega/(2\pi)=10$ GHz in panels (d, e).}
\label{figterms}
\end{figure*}

As the nanowire is embedded in other materials, we assume that the phonons are weakly confined in silicon and thus make use of the 3D model [Eq. (\ref{eq3Drelax})] for the calculation of the relaxation time. The actual impact of dimensionality and encapsulation materials will be investigated in sections \ref{sectionbulklimit} and \ref{sectionembedding}. Eq. (\ref{eq3Drelax}) contains many different terms that can be categorized as a function of {\it i}) their nature [$A_n$ (dipole) or $B_n$ (homogeneous band mixing) terms]; {\it ii}) the polarization of the phonons (LA or TA) and {\it iii}) the deformation potentials involved. In order to sort Eq. (\ref{eq3Drelax}) into these categories, the relaxation rate $\Gamma_{\rm ph}^{\rm 3D}$ is plotted as a function of the Larmor frequency $\omega$ of the qubit in Figs.~\ref{figterms}a-c ($V_{\rm bg}=-0.15$ V) and decomposed (a) into $A_n$ and $B_n$ contributions, (b) LA and TA contributions, and (c) $a$, $b$ and $d$ contributions (the other deformation potentials being set to 0). It is clear from Fig.~\ref{figterms} that the band mixing $B_n$ terms dominate the relaxation through the coupling to TA phonons by the shear deformation potential $d$. The $B_2\Lambda^{\rm B}_{2t}\propto d^2$ term of Eq. (\ref{eq3Drelax}) actually makes the largest contribution to the relaxation rate, followed by the $B_1\Lambda^{\rm B}_{1t}\propto b^2$ term. This implies that we can make the approximation $e^{i\vec{q}\cdot\vec{r}}\sim 1$ in Eq. (\ref{eqgoldenrule2}) and drop all dipole $A_n$ terms arising from the {\CORR first-order} expansion of this phase factor. This is opposite to electrons in the single band effective mass approximation, whose relaxation is exclusively ruled by such dipole terms. The action of spin-orbit coupling within the valence band is, indeed, very dependent on the heavy- and light-hole balance and is, therefore, sensitive to band mixing by phonons. This will be further discussed in section \ref{sectionoptimal}. We also emphasize that the dipole terms can be dominant for charge relaxation {\CORR between states whose main envelopes are orthogonal}.

The relaxation rate is $\propto\omega^4$ at low Larmor frequency and $\propto\omega^5$ at large Larmor frequency. The $\propto\omega^5$ behavior results from the $\propto\omega^3$ prefactor of the $B_n$ terms (phonon strains and density of states) and from the $\propto\omega^2$ dependence of the $B_n$ terms themselves. Indeed, as discussed in section \ref{section1D2D}, the $B_n$ terms vanish at zero magnetic field when $\ket{\zero}$ and $\ket{\one}$ are time-reversal symmetrics one of each other, and increase as $\omega^2$ once a finite magnetic field breaks time-reversal symmetry. Accordingly, the (however negligible) $A_n$ term show a $\propto\omega^7$ behavior (owing to the additional $q^2\propto\omega^2$ dependence of the dipole terms). At small magnetic field, the relaxation rate departs from the $\propto\omega^5$ behavior due to the $\coth[\hbar\omega/(2k_BT)]$ prefactor that accounts for the larger population of acoustic phonons at the Larmor frequency (enhancement of absorption and stimulated emission processes). 

The contributions of LA and TA phonons to the relaxation rate are plotted as a function of back gate voltage in Figs. \ref{figterms}d,e. The conclusions drawn above remain valid over the whole range of investigated back gate voltages. We discuss in more detail the dependence of $\Gamma_{\rm ph}^{\rm 3D}$ on $V_{\rm bg}$ in the next section.

\subsection{Optimal operation point}
\label{sectionoptimal}

\begin{figure*}
\includegraphics[width=1.8\columnwidth]{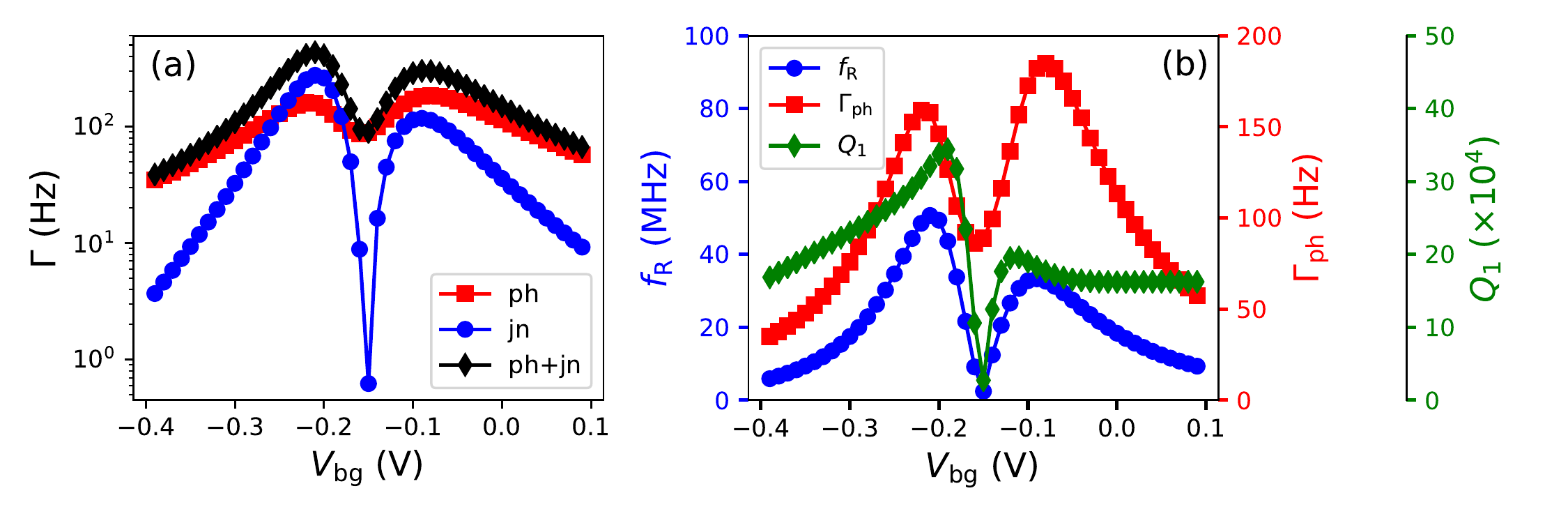}
\includegraphics[width=1.8\columnwidth]{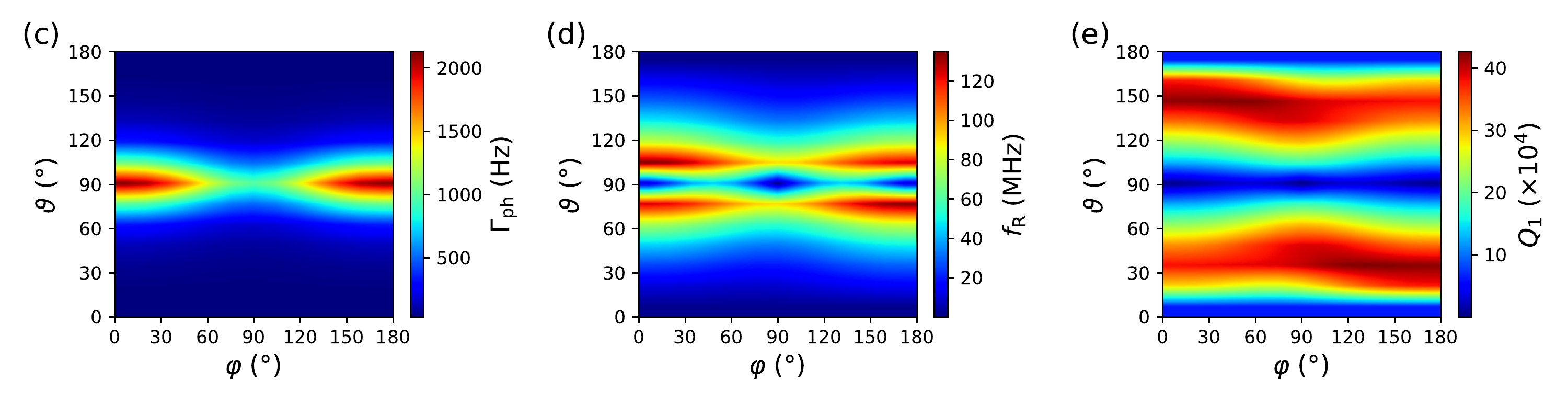}
\caption{(a) Johnson-Nyquist ($\Gamma_{\rm jn}$), phonon-induced ($\Gamma_{\rm ph}^{\rm 3D}$) and total relaxation rate as a function of back gate voltage $V_{\rm bg}$. $\Gamma_{\rm jn}$ is computed for a resistance $R=200$ $\Omega$ on the central front gate. (b) Rabi frequency $f_{\rm R}$, phonon-induced relaxation rate $\Gamma_{\rm ph}^{\rm 3D}$, and quality factor $Q_1=f_{\rm R}/\Gamma_{\rm ph}^{\rm 3D}$ as a function of $V_{\rm bg}$. (c) Relaxation rate $\Gamma_{\rm ph}^{\rm 3D}$, (d) Rabi frequency $f_{\rm R}$, and (e) quality factor $Q_1=f_{\rm R}/\Gamma_{\rm ph}^{\rm 3D}$ as function of the orientation of the magnetic field characterized by the azimuthal and polar angles $\theta$ and $\varphi$ defined on Fig. \ref{figdevice} (and consistent with Refs. \onlinecite{Crippa18}, \onlinecite{Venitucci18} and \onlinecite{Venitucci19}), at $V_{\rm bg}=-0.2$ V. The Larmor frequency is $\omega/(2\pi)=10$ GHz in all plots. The orientation of the magnetic field is $\theta=45^\circ$, $\varphi=0$ in panels (a) and (b). Note that panel (d) looks different from  Ref. \onlinecite{Venitucci18} because it is plotted at constant Larmor frequency instead of constant magnetic field amplitude.}
\label{figoptimal}
\end{figure*}

The total relaxation rate is plotted as a function of $V_{\rm bg}$ in Fig. \ref{figoptimal}a. We have added for comparison the relaxation rate $\Gamma_{\rm jn}$ due to Johnson-Nyquist noise on the central front gate (zero-point and thermal fluctuations in the circuit connected to that gate):\cite{Hu14,Clerk10}
\begin{equation}
\Gamma_{\rm jn}=4\pi\frac{R}{R_0}\Big|\langle\zero|D_{\rm fg}|\one\rangle\Big|^2\omega{\rm coth}\left(\frac{\hbar\omega}{2k_BT}\right)\,, 
\end{equation}
where $R$ is the resistance connected to the gate, $R_0=h/e^2$, and $D_{\rm fg}(\vec{r})=\partial V_t(\vec{r})/\partial V_{\rm fg}$ is the derivative of the total potential $V_t(\vec{r})$ in the device with respect to the front gate voltage $V_{\rm fg}$. We assume $R=200$ $\Omega$.

The relaxation rate $\Gamma_{\rm ph}^{\rm 3D}$ is plotted along with the Rabi frequency $f_{\rm R}$ in Fig. \ref{figoptimal}b. The Rabi oscillations are driven by a radio-frequency modulation $\delta V_{\rm fg}=1$ mV on the front gate. The quality factor $Q_1=f_{\rm R}/\Gamma_{\rm ph}^{\rm 3D}$ is also plotted on that figure. It gives the number of Rabi oscillations that can be achieved within one relaxation time $T_1=\Gamma_{\rm ph}^{-1}$. We have only accounted for phonons in this figure, which therefore provides an upper limit to the quality factor of the qubit. Other mechanisms for relaxation (Johnson-Nyquist and charge noise...) are indeed extrinsic to the qubit and are, in principle, more amenable to optimization (e.g., by the reduction of circuit impedances for Johnson-Nyquist noise). Multi-phonon and photon processes (relevant at high enough temperature),\cite{Loss09,Petit18} and additional mechanisms for spin-orbit coupling neglected in this study (through remote coupling to the conduction bands in particular\cite{Winkler03}) may also degrade this figure of merit. The impact of Johnson-Nyquist noise on the quality factor and coherence time is discussed in Appendix \ref{appendixJN}.

The relaxation rate shows modulations as a function of $V_{\rm bg}$ that mimic those of the Rabi frequency. In particular, $\Gamma_{\rm ph}^{\rm 3D}$ displays a dip near $V_{\rm bg}=-0.15$ V where the hole wave functions feature an approximate inversion center.\cite{Venitucci18} This hampers the action of spin-orbit coupling and decouples the holes from the radio-frequency electric field from the front gate. The Rabi oscillations are therefore slow but the holes get also decoupled from Johnson-Nyquist and charge noise at that point. 

The dip in $\Gamma_{\rm ph}^{\rm 3D}$ is not, however, as marked as the dip in the Rabi frequency and $\Gamma_{\rm jn}$. This follows, in particular, from the fact that the band mixing terms can not be cast as the action of an effective electric field due to phonons; only the hydrostatic $\propto a$ terms can be so. The decrease of the relaxation rate at large positive or negative back gate voltage is due to the strong lateral confinement in the static electric field of the gate.\cite{Venitucci18,Venitucci19} Although the lifetime is longer at large $|V_{\rm bg}|$, the Rabi frequency is smaller, which slightly lowers the quality factor. 

The phonon-limited lifetimes are typically shorter (yet still $>5$ ms) than expected in electron qubits\cite{Tahan14,Hu14,Bourdet18} owing to the strong spin-orbit coupling in the valence band. This is however balanced by much larger Rabi frequencies, allowing for significant quality factors. Hole qubits are also much more sensitive to Johnson-Nyquist noise than electron qubits.\cite{Hu14,Bourdet18} As a matter of fact, electrical and charge noise is presumably dominating decoherence and relaxation in present hole qubit devices.\cite{Maurand16} The phonon-limited quality factor is maximal near $V_{\rm bg}=-0.2$ V. As suggested by Fig. \ref{figoptimal}a, the position of the optimal bias point for manipulation will move further away from $V_{\rm bg}=-0.15$ V when increasing electrical and charge noise (see Appendix \ref{appendixJN}). Anyhow, the qubit may be brought back to the ``sweet spot'' $V_{\rm bg}=-0.15$ V in between manipulations, where the lifetime is longest.\cite{Kloeffel13,Maier13}

The Rabi frequency, relaxation rate $\Gamma_{\rm ph}^{\rm 3D}$ and quality factor are also strongly dependent on the magnetic field orientation (Figs. \ref{figoptimal}c,d,e). The orientational dependence of $f_R$ has been discussed in detail in Refs. \onlinecite{Venitucci18} and \onlinecite{Venitucci19} and is the fingerprint of the mostly heavy-hole character of the qubit states and of the symmetries of the device. Following the lines of Ref. \onlinecite{Venitucci19}, we reach the following expressions for the dominant $B_1$ and $B_2$ terms [Eqs. (\ref{eqBs})] near the ``sweet spot'' $V_{\rm bg}=-0.15$ V, at the leading order in the channel height $H$:
\begin{subequations}
\label{eqB12s}
\begin{align}
B_1&=\frac{m_0^2H^4}{2\hbar^2\gamma_2^2}(\kappa\mu_B B)^2\sin^2\theta \\
B_2&=2B_1\,,
\end{align}
\end{subequations}
where $m_0$ is the free electron mass, $\gamma_2=0.339$ is a valence band Luttinger parameter, $\kappa=-0.42$ is the Zeeman coefficient of the holes and $\mu_B$ is Bohr's magneton.\cite{Winkler03} $B_1$ and $B_2$ behave, as expected, as $\omega^2\propto B^2$. At this order in $H$, they do not depend on the width $W$ of the channel, hence on lateral confinement. {\CORR In this respect, phonon-induced relaxation behaves differently than Johnson-Nyquist relaxation and Rabi oscillations, which do require lateral confinement.\cite{Venitucci19} This results from the fact that strains can couple directly heavy- and light-hole Bloch functions, at variance with a radio-frequency electric field.} Actually, the $\propto B\sin\theta$ in-plane magnetic field mixes the majority, heavy $|3/2,+3/2\rangle$ component of $\ket{\one}$ with a light $|3/2,+1/2\rangle$ envelope, which can then be coupled by the phonons to the majority $|3/2,-3/2\rangle$ component of $\ket{\zero}$ through the strain Hamiltonian $\Delta H(\varepsilon)$ [Eq. (\ref{eqdeltaH})]. The magnetic mixing between the $|3/2,+3/2\rangle$ and $|3/2,+1/2\rangle$ envelopes of $\ket{\one}$ is inversely proportional to the splitting between the confined heavy- and light-hole subbands, which gives rise to the $H^4/\gamma_2^2$ dependence in Eqs. (\ref{eqB12s}) {\CORR (as in a quantum well)}. The relaxation rate is hence maximal near $\theta=\pi/2$; the dependence on $\theta$ on Fig. \ref{figoptimal}c is even strenghtened because the effective $g$ factor of heavy-holes is minimal at $\theta=\pi/2$, so that larger magnetic fields are needed to reach the target Larmor frequency $\omega/(2\pi)=10$ GHz. The quality factor is weakly dependent on $\varphi$ and is significant in a wide band of $\theta$'s. It peaks near $\theta=30^\circ$ and $\theta=150^\circ$, close to the reference orientation chosen in Figs. \ref{figoptimal}a,b. Note that $f_R\propto\sin\theta$ when $\theta\to0$ or $\pi$ so that Eqs. (\ref{eqB12s}) suggest that $Q_1\to\infty$ (although this is irrelevant since $f_R\to0$). Higher-order contributions to Eq. (\ref{eqB12s}) give rise to a finite $\Gamma_{\rm ph}$ and to the dependence on $\varphi$ clearly visible on Fig. \ref{figoptimal}c, ensuring that $Q_1\to0$ when $\theta\to0$ or $\pi$ as shown in Fig. \ref{figoptimal}e.

\section{Effects of phonon confinement and encapsulation materials}
\label{sectionconfinement}

In this section, we discuss the effects of phonon confinement\cite{Kargar16} on the relaxation rate, as well as the impact of encapsulation materials. We highlight how the lifetime of the qubit depends on its vibrational environment over long length scales.

\subsection{Phonon confinement}
\label{sectionbulklimit}

\begin{figure}
\includegraphics[width=0.9\columnwidth]{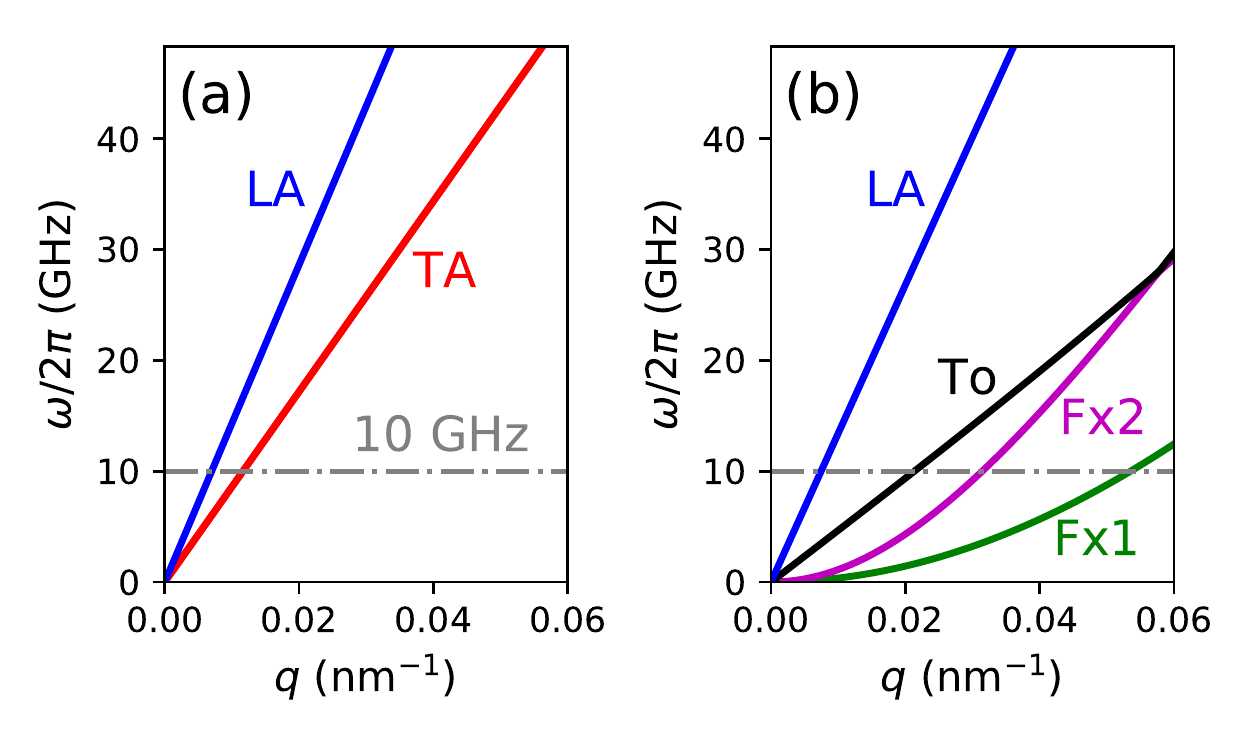}
\includegraphics[width=0.9\columnwidth]{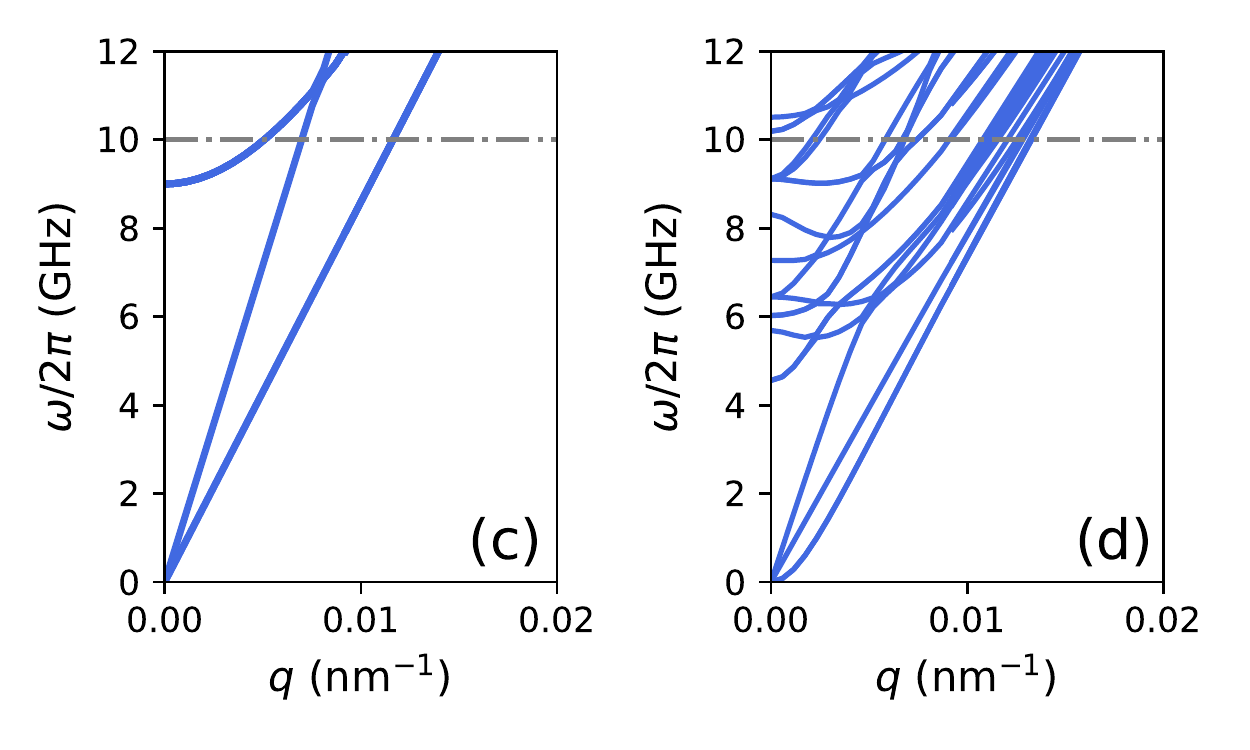}
\caption{Phonon band structures of a $l_y=30{\rm\ nm}\times l_z=10{\rm\ nm}$ silicon nanowire with (a) periodic boundary conditions and (b) free-standing boundary conditions; and of a $l_y=600{\rm\ nm}\times l_z=600{\rm\ nm}$ silicon wire with (c) periodic boundary conditions and (d) free-standing boundary conditions. The isotropic elastic constants of silicon were used in this calculation (table \ref{tablecij}). The gray dash-dotted line is the Larmor frequency of the qubit $\omega/(2\pi)=10$ GHz. The linear torsional (To) and parabolic flexural (Fx1 and Fx2) modes of the free-standing nanowire are clearly visible in panel (b).}
\label{figBS1}
\end{figure}

In section \ref{sectiontheory}, we have derived the relaxation rate for bulk phonons [Eq. (\ref{eq3Drelax})], strongly confined 2D [Eq. (\ref{eq2Drelax})] and strongly confined 1D [Eq. (\ref{eq1Drelax})] phonons. In this section, we validate the 3D, 2D and 1D expressions on numerical calculations of the phonon band structure and address their range of validity. We discuss the impact of phonon confinement on the relaxation rate of the qubit.

For that purpose, we consider the same qubit as in Fig. \ref{figdevice}, but coupled to the phonon band structure of a square $[110]$-oriented wire with varying side $l_y=l_z$, or of a rectangular $[110]$-oriented wire with side $l_z=H=10$ nm and varying $l_y$. In the first case (square wire), we expect a transition from a 1D regime at small $l_y=l_z$ to a 3D regime at large $l_y=l_z$, and in the second case (rectangular wire) a transition from a 1D regime at small $l_y$ to a 2D regime at large $l_y$. The relaxation rates are computed with Eq. (\ref{eqGammanum}).

We consider either {\it i}) free-standing boundary conditions (no stress perpendicular to the surfaces) or {\it ii}) periodic Born-von-Karman boundary conditions at the surface of the wires. For periodic boundary conditions, the resulting phonon band structure is nothing else than the bulk band structure sampled at wave vectors $\vec{q}=(q, 2\pi n_y/l_y, 2\pi n_z/l_z)$ (in the device axes frame), where $q$ is the 1D wave vector of the wire and $n_y$, $n_z$ are integers. Each pair $(n_y, n_z)$ defines a set of three sub-bands (sampled in the bulk LA, TA1 and TA2 branches). The acoustic branches ($\omega_{\alpha q}\to0$ when $q\to0$) of the wire are the $n_y=n_z=0$ sub-bands. In this approximation, the displacements are homogeneous in the cross-section of the wire. The other 1D sub-bands have finite $\omega_{\alpha 0}$. In the strongly confined regime (only the three acoustic branches of the wires couple to the qubit), the relaxation rate shall therefore exactly match the expressions for $\Gamma_{\rm ph}^{\rm 1D}$ or $\Gamma_{\rm ph}^{\rm 2D}$ that were established under these assumptions. On the opposite, in the weakly confined regime (many sub-bands couple to the qubit), the relaxation rate shall tend to $\Gamma_{\rm ph}^{\rm 3D}$. This provides a numerical test of these expressions, and allows for a clear investigation of the transition from one regime to an other. Free-standing boundary conditions are more relevant for truly confined phonons with inhomogeneous displacements in the wire cross section. Also, the phonon band structure of a free-standing wire shows specific features:\cite{Nishiguchi97,Thonhauser04} there are, in particular, four branches whose $\omega_{\alpha q}\to0$ when $q\to0$ (one linear longitudinal and one linear torsional mode, and two parabolic flexural modes). This reflects the translational and rotational invariances of the elastic energy of a free-standing structure. We will discuss the contributions of each mode to the relaxation rate and the relevance of the 1D formula in this context.

\begin{figure*}
\includegraphics[width=1.8\columnwidth]{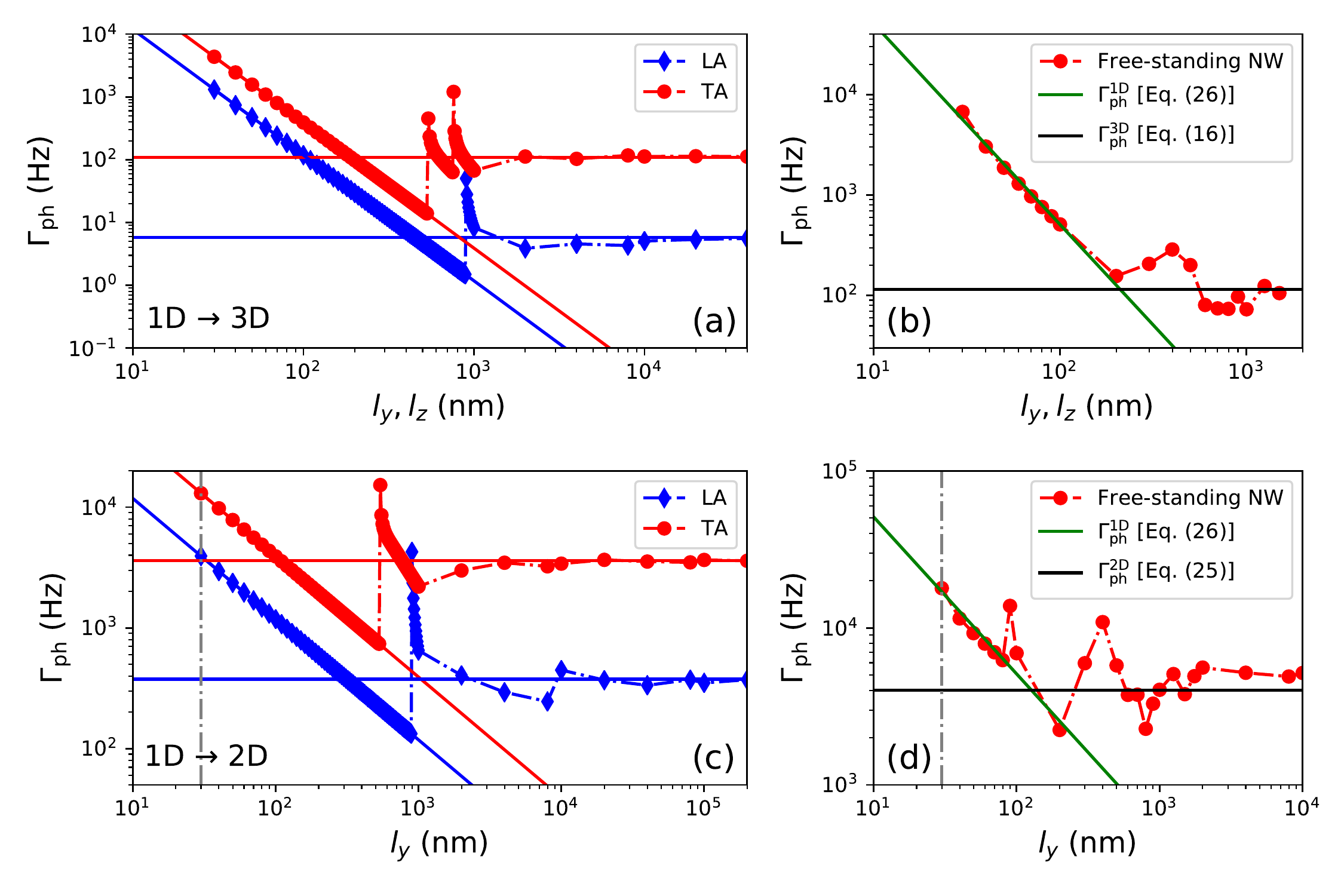}
\caption{Relaxation rate $\Gamma_{\rm ph}$ computed in a qubit embedded (from a vibrational point of view) in (a) a square nanowire (NW) with sides $l_y=l_z$ and periodic boundary conditions, (b) the same square nanowire with free-standing boundary conditions, (c) a rectangular nanowire with side $l_z=10$ nm, varying $l_y$, and periodic boundary conditions, and (d) the same rectangular nanowire with free-standing boundary conditions. For periodic boundary conditions, the phonon band structure is sampled from the bulk LA and TA branches and the corresponding contributions are displayed separately. The semi-analytical results for $\Gamma_{\rm ph}^{\rm 1D}$, $\Gamma_{\rm ph}^{\rm 3D}$ (panels a, b) and $\Gamma_{\rm ph}^{\rm 2D}$ (panels c, d) are also plotted as solid lines. The transitions from the 1D to 3D regimes (panels a, b) and from the 1D to 2D regimes (panels c, d) are clearly visible. All data were computed with the isotropic model for the elastic constants of silicon (see Table \ref{tablecij}). The back gate voltage is $V_{\rm bg}=-0.2$ V, and the orientation of the magnetic field is $\varphi=\theta=45^\circ$.}
\label{figdim}
\end{figure*}

The phonon band structure of a $l_y=30{\rm\ nm}\times l_z=10{\rm\ nm}$ wire is plotted as an illustration in Figs. \ref{figBS1}a,b, for both periodic and free-standing boundary conditions. We use the isotropic model for the elastic constants of Silicon (Table \ref{tablecij}). In the free-standing case, the phonon band structure is computed with the numerical finite-differences approach outlined in section \ref{sectionnumerical}. The two parabolic flexural modes, as well as the torsional and longitudinal branches are clearly visible on this figure. The phonon band structures of a $l_y=600{\rm\ nm}\times l_z=600{\rm\ nm}$ wire with free-standing and periodic boundary conditions are likewise plotted in Figs. \ref{figBS1}c,d. The energy and wave vector range where the flexural branches are parabolic decreases with increasing wire size and the acoustic group velocities outside this range get closer to those of the bulk LA and TA phonons. The number of sub-bands below the Larmor frequency is larger with free-standing than with periodic boundary conditions (with the appearance of, e.g., breathing modes), but the sub-bands are much more degenerate in the latter case.

For periodic boundary conditions, the relaxation rate in the square wires is plotted as a function of $l_y=l_z$ in Fig. \ref{figdim}a, and the relaxation rate in the rectangular wires is plotted as a function of $l_y$ in Fig. \ref{figdim}c. Both LA and TA contributions are displayed. The numerical data are compared with the models for the 1D and 2D or 3D phonon band structures.

First of all, the transitions from the 1D to 3D regime, and from the 1D to the 2D regime are clearly visible in these figures. As expected, the analytical expressions of section \ref{sectiontheory} reproduce very well the numerical data in these different regimes. In the 1D regime, the relaxation rate is inversely proportional to the cross-sectional area $S=l_yl_z$ of the wire [Eq. (\ref{eq1Drelax})], as the overlap between the squared acoustic phonon and qubit wave functions scales as $1/S$. The relaxation rate in the 1D regime can hence be much larger than in the 3D regime depending on $S$ (and on the Larmor frequency $\omega$). The transition from the 1D to the 3D regime occurs in the range $l_y=l_z\simeq 1000$ nm -- {\CORR which is comparable to the typical wave length of the 3D phonons involved in the relaxation but} very large with respect to the size of the qubit itself. This results from the fact that the splitting between the 1D phonon sub-bands remains greater that the Larmor frequency ($10$ GHz) over a wide range of dimensions, so that only the 1D LA and TA branches can contribute to scattering until $l_y=l_z\simeq 500$ nm. The introduction of a new sub-band gives rise to a peak in the relaxation rate (due to the Van-Hove singularity in the density of states of a parabolic sub-band), until the number $N_{\rm ph}$ of sub-bands at the Larmor frequency is large enough to reach the 3D limit. Indeed, at wide enough $l_y=l_z$, $N_{\rm ph}\gg 1$ is approximately given by:
\begin{equation}
N_{\rm ph}\simeq \frac{S}{(2 \pi)^2}\sum_{\alpha\in\{l,t_1,t_2\}}\pi\left(\frac{\omega}{v_\alpha}\right)^2\,,
\label{eqNph}
\end{equation}
where $v_l$, $v_{t_1}=v_{t_2}=v_t$ are the longitudinal and transverse sound velocities in bulk silicon. The $\propto S$ dependence of $N_{\rm ph}$ balances the $\propto 1/S$ dependence of the squared phonon amplitudes, so that the relaxation rate becomes independent on $S$ (see Appendix \ref{appendixConvergence} for a discussion on the nature of the convergence). The same conclusions hold for the 1D to 2D transition, which also occurs around $l_y=1000$ nm. 

The relaxation rates are plotted for free-standing boundary conditions in Fig. \ref{figdim}b,d. The equations for $\Gamma_{\rm ph}^{\rm 1D}$ and $\Gamma_{\rm ph}^{\rm 2D}$ still hold surprisingly well in the strongly confined regime despite the differences between periodic and free-standing phonon band structures (Fig. \ref{figBS1}). The transition from the 1D to 2D or 3D regime takes place at significantly smaller lateral dimensions ($\simeq 500$ nm) due to the larger density of low energy sub-bands. In the 1D limit, the qubit mostly couples to the longitudinal and flexural modes of the wire. We emphasize, though, that the qubit was placed near the highly symmetric central position of the square or rectangular wire where the effects of boundary conditions are expected to be minimal. 

Irrespective of the choice of boundary conditions, Fig. \ref{figdim} clearly highlights the effect of phonon confinement on the relaxation rate in the qubit. $\Gamma_{\rm ph}$ can be strongly dependent on the geometry far away from the qubit as the spin couples to very long wave length energy acoustic phonons that can probe the device over hundreds of nanometers. We further support this conclusion in the next section by studying the impact of the encapsulation materials on the relaxation in the qubit.

\subsection{Impact of encapsulation materials}
\label{sectionembedding}

\begin{figure*}
\includegraphics[width=1.8\columnwidth]{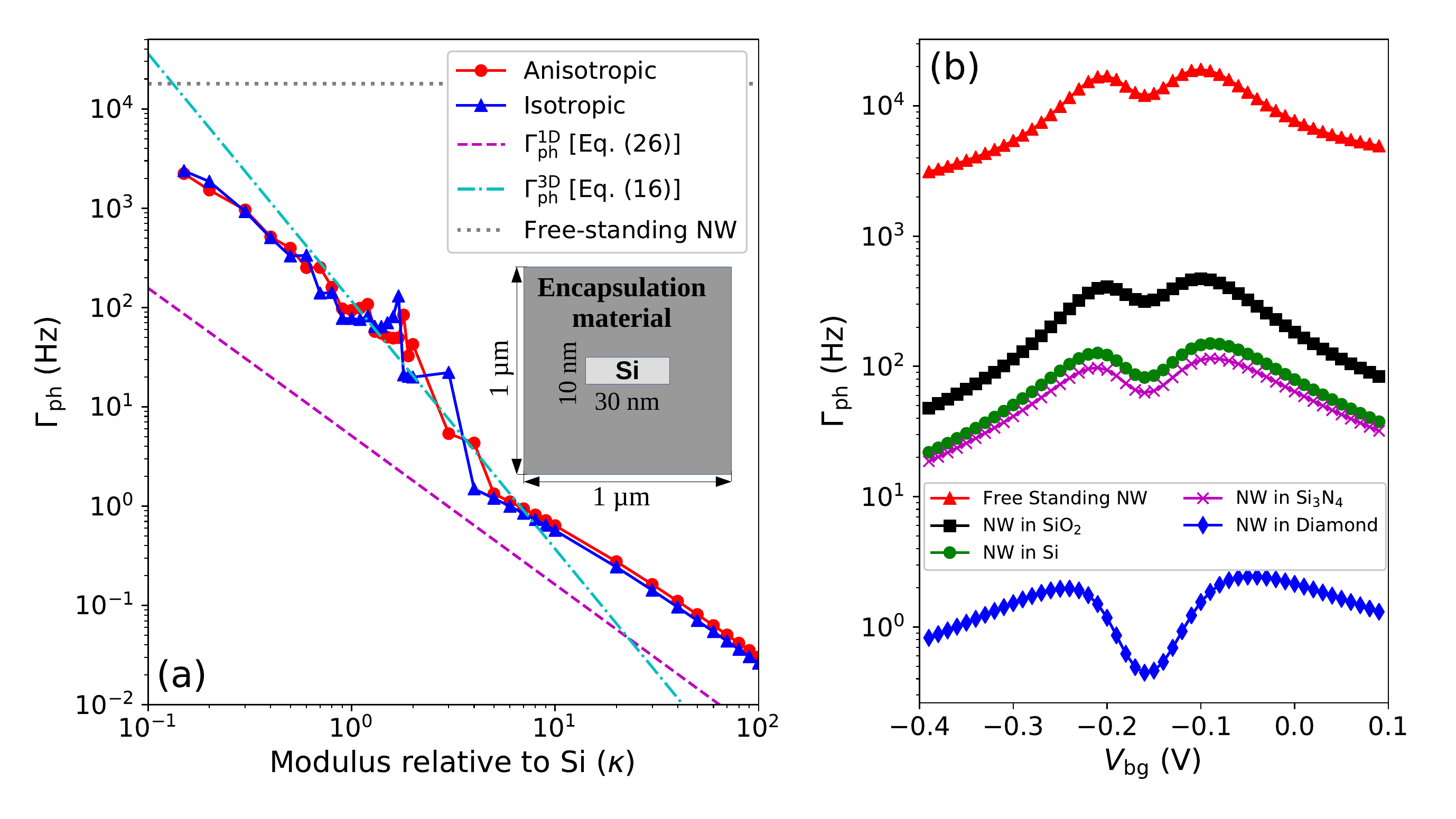}
\caption{(a) Phonon-induced relaxation rate as a function of the hardness of the encapsulation material. From a vibrational point of view, the system is modeled as a $W=30{\rm\ nm}\times H=10{\rm\ nm}$ silicon nanowire (NW) embedded in a material whose elastic constants are rescaled by a factor $\kappa$ with respect to those of silicon. Periodic boundary conditions are applied over a supercell with side $L_{\rm cell}=1\ \mu$m. Data are plotted for both isotropic and anisotropic elastic models (see Table \ref{tablecij}). The free-standing limit, as well as the semi-analytical results for $\Gamma_{\rm ph}^{\rm 3D}$ and $\Gamma_{\rm ph}^{\rm 1D}$ (with rescaled phonon velocities, see main text) are displayed for comparison. The back gate voltage is $V_{\rm bg}=-0.2$ V, and the orientation of the magnetic field is $\varphi=\theta=45^\circ$. (b) Phonon-induced relaxation rate as a function of back gate voltage for different encapsulation materials, that are either soft (free-standing limit, SiO$_2$), comparable (Si, Si$_3$N$_4$), or harder (diamond) than silicon ($\theta=45^\circ$, $\varphi=0$). The Larmor frequency is $\omega=10$ GHz in all panels.}
\label{figenca}
\end{figure*}

As discussed previously, the qubits are usually embedded in complex stacks of materials, which can have a significant impact on the long-wave length acoustic phonons that couple to the spins. In particular, the velocity and degree of confinement of the phonons is highly dependent on the hardness of the materials around the channel. In order to explore this issue, we compute the phonon band structure of the device of Fig. \ref{figdevice} modeled as a rectangular silicon nanowire with sides $W=30$ nm and $H=10$ nm embedded in a homogeneous material with varying elastic constants. We apply periodic boundary conditions over the cross-section in a supercell with sides $l_y=l_z=L_{\rm cell}=1\ \mu$m.

We use both the standard (anisotropic) elastic constants of silicon and the isotropic modification giving rise (in bulk) to a LA branch with velocity $v_t=9000$ m/s and to two degenerate TA branches with velocity $v_t=5400$ m/s, consistent with the analytical 3D model of section \ref{sectiontheory} (see Table \ref{tablecij}). The elastic constants of the encapsulation material are rescaled by a factor $\kappa$ with respect to those of silicon but the density is the same.

The relaxation rate $\Gamma_{\rm ph}$ is plotted as a function of $\kappa$ in Fig. \ref{figenca}a. The horizontal dotted line is the relaxation rate computed in a free-standing $W=30{\rm\ nm}\times H=10{\rm\ nm}$ nanowire. As expected, $\Gamma_{\rm ph}$ tends to this limit when $\kappa\to 0$ (encapsulation material much softer than silicon). The relaxation rate then decreases continuously as the encapsulation material becomes harder and harder. The results obtained with the isotropic and anisotropic models for the elastic constants are very close, which shows that the moderate anisotropy of silicon does not have much impact on the relaxation.

In order to get further insights into these trends, we plot in Figs. \ref{figBS2}a,b,c the 1D phonon sub-band structure computed at $\kappa=0.1$, $\kappa=1$ and $\kappa=10$ (isotropic model). The horizontal dash-dotted line on these plots is the Larmor frequency of the qubit $\omega/(2\pi)=10$ GHz. The longitudinal and transverse sound velocities (drawn from the three branches whose $\omega_{\alpha q}\to0$ when $q\to0$) are plotted as a function of $\kappa$ in Figs. \ref{figBS2}d, and the number of 1D sub-bands that cross the Larmor frequency of the qubit is plotted in Fig. \ref{figBS2}e.

\begin{figure*}
\includegraphics[width=1.8\columnwidth]{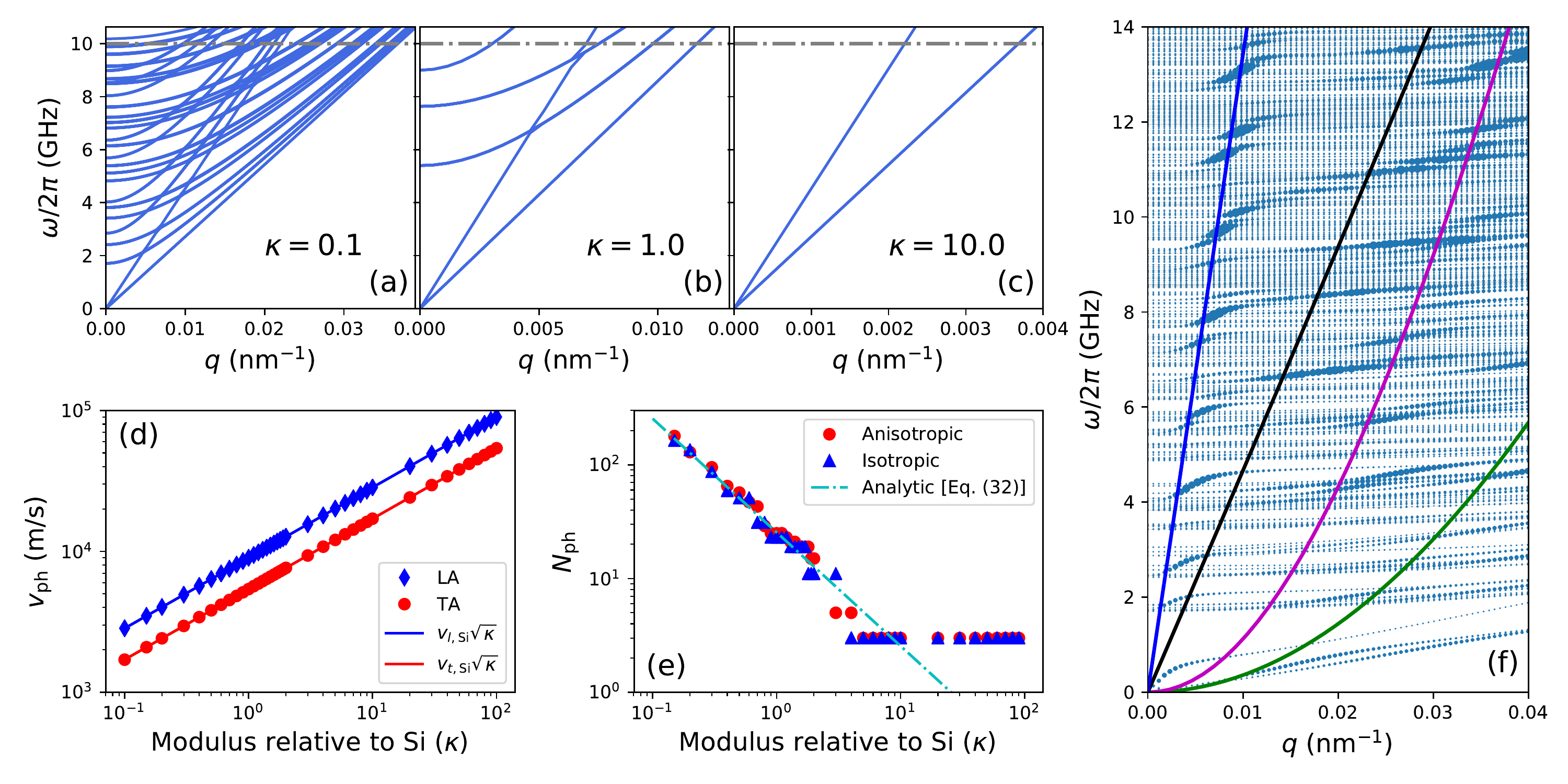}
\caption{(a, b, c) Phonon band structures computed for $\kappa=0.1$, $\kappa=1$ and $\kappa=10$ (isotropic elastic constants). (d) The transverse and longitudinal acoustic velocities (drawn from the branches $\omega_{\alpha q}\to0$ when $q\to0$) as a function of $\kappa$. The group velocities are defined by the encapsulation material that fills $99.97\%$ of the supercell and behave, therefore, as $v_{\rm enc}=\sqrt{\kappa}v_{\rm Si}$. (e) The number of phonon branches $N_{\rm ph}$ that match the Larmor frequency $\omega/(2\pi)=10$ GHz of the qubit. The dash-dotted line is Eq. (\ref{eqNph}) using the sound velocities $v_{\rm enc}$ as input. (f) The phonon band structure computed at $\kappa=0.001$ and $L_{\rm cell}=100$ nm ($L_{\rm cell}$ being reduced here due to computational limitations). The diameter of the dots is proportional to the density of elastic energy $U_{\rm el}$ in the silicon core.\cite{notedefU} The solid color lines  are the longitudinal, torsional and flexural modes of a free-standing $W=30{\rm\ nm}\times H=10{\rm\ nm}$ silicon nanowire. At such small $\kappa$, most of the low-energy phonons propagate in the encapsulation material (phonon depletion effect) and do not couple efficiently to the qubit; yet the phonon modes of the free-standing nanowire emerge behind the quasi-continuum of states of the encapsulation material.}
\label{figBS2}
\end{figure*}

The sound velocities $v\simeq v_{\rm enc}=\sqrt{\kappa}v_{\rm Si}$ are defined by the encapsulation material that fills $99.97\%$ of the supercell ($v_{\rm Si}$ and $v_{\rm enc}$ being respectively the sound velocities in silicon and in the encapsulation material). Accordingly, the density of 1D phonon sub-bands is essentially proportional to $1/v_{\rm enc}^2$, so that the number $N_{\rm ph}$ of phonon sub-bands at the Larmor frequency is given by Eq. (\ref{eqNph}) with $v_\alpha$ replaced by $v_{\rm enc,\alpha}$ {\CORR (and $S$ by $L_{\rm cell}^2$)}. $N_{\rm ph}$ therefore behaves as $1/\kappa$. This trend is clearly visible in Figs. \ref{figBS2}a,b,c: the number of phonon sub-bands that cross the Larmor frequency increases with decreasing $\kappa$. For $\kappa=1$, the phonon wave functions are simply the bulk wave functions delocalized over the whole supercell. As a consequence, the relaxation rate scales as $\Gamma_{\rm ph}^{\rm 3D}\propto 1/v_{\rm enc}^5\propto\kappa^{-5/2}$ near $\kappa=1$ [Eq. (\ref{eq3Drelax}) using $v_{\rm enc}$ as input]. When decreasing $\kappa\ll1$, $N_{\rm ph}$ increases continuously; yet all the sub-bands do not contribute equally to the relaxation rate. Indeed, many low-energy sub-bands mostly propagate in the encapsulation material at small $\kappa$ and are, therefore, weakly coupled to the qubit (phonon depletion effect\cite{Pokatilov04,Pokatilov05,Pokatilov05b}). When $\kappa\to0$, the LA and two TA branches of the silicon nanowire actually emerge behind the quasi-continuum of sub-bands propagating in the encapsulation material, and are the only ones that efficiently scatter the hole (see Fig. \ref{figBS2}f). The relaxation rate hence saturates at the free-standing limit. On the contrary, when the encapsulation material is hardened ($\kappa\gg1$), the splitting between phonon sub-bands increases until the qubit can only couple to the LA and two TA sub-bands whose $\omega_{\alpha q}\to0$ when $q\to0$ (Fig. \ref{figBS2}c). These three sub-bands are highly dispersive (with group velocities that scale as $\sqrt{\kappa}$), and tend to relocalize in the silicon core at finite $q$.\cite{notephonons} In this regime, $\Gamma_{\rm ph}$ behaves as $1/(v_{\rm enc}^3L_{\rm cell}^2)\propto\kappa^{-3/2}$, as expected from Eq. (\ref{eq1Drelax}) for the 1D phonons model. The relaxation rate remains, however, significantly larger than Eq. (\ref{eq1Drelax}) due to the finite phonon reconfinement in the silicon core at small wave vector $q\propto\sqrt{\kappa}$.\cite{notephonons} The residual relaxation rate also becomes dependent on $L_{\rm cell}$, being sensitive to details of the structure over the scale of the phonon wave length.\cite{notephonons2} 

In Fig.~\ref{figenca}b, we compare the relaxation rates computed in Si nanowires embedded in different materials, as a function of the back gate voltage. The elastic constant of these materials are given in Table \ref{tablecij}. They are not all meant to be realistic encapsulation materials for a silicon nanowire qubit, but have been chosen as representatives of ``soft'' and ``hard'' materials. We have also neglected built-in strains as well as disorder and (if relevant) piezo-electric scattering in these materials. We recover the trends discussed above: the softer the encapsulation material, the shorter the phonon-limited lifetime. In particular, the bulk relaxation rate $\Gamma_{\rm ph}^{\rm 3D}$ [Eq. (\ref{eq3Drelax})] typically overestimates the lifetime as the encapsulation materials (such as SiO$_2$) are usually softer than silicon. Phonon engineering in semiconductor qubits may, therefore, ultimately improve their performances, once all other extrinsic sources of scattering have been mitigated.\cite{Balandin12}

\section{Conclusions}

We have derived the phonon-limited lifetime in hole spin-orbit qubits within the 6 bands $\vec{k}\cdot\vec{p}$ framework, accounting for the complete set of deformation potentials of the valence band. The resulting expressions for the {\CORRR one-phonon} transition rates can actually be applied to both spin and charge relaxation in a hole quantum dot. We have extended these expressions to strongly confined 1D and 2D phonon band structures and highlighted the different dependences on the Larmor frequency of the qubit. We have then applied this theory to a hole spin-orbit qubit on silicon-on-insulator similar to Refs. \onlinecite{Crippa18} and \onlinecite{Venitucci18}. We have shown that phonon-induced spin relaxation in this qubit is dominated by a band mixing term that couples the hole to transverse acoustic phonons through the valence band deformation potential $d$. We have next optimized the bias point and magnetic field orientation looking for the best quality factor $Q_1=f_RT_1$ (the number of Rabi oscillations that can be performed within one relaxation time $T_1$). When only phonons are accounted for in the relaxation, $Q_1$ can reach a few tens of thousands despite the strong spin-orbit coupling in the valence band. Hole spin-orbit qubits are, however, very sensitive to electrical and charge noise, which calls for a careful design of the devices and of the electronics around. We have also discussed the impact of confinement and encapsulation materials on the phonon-limited lifetimes. Indeed, the qubit couples to low-energy phonons that probe the device over very long length scales. The lifetime does, in particular, increase when the materials around the qubit get harder. This may be evidenced experimentally at magnetic fields large enough so that phonons dominate over electrical and charge noise relaxation. Phonon engineering might, therefore, ultimately improve the performances of semiconductor qubits.

\begin{acknowledgments}
This work was supported by the European Union's Horizon 2020 research and innovation program under grant agreements No 688539 MOSQUITO and 810504-QUCUBE-ERC-2018-SyG, and by the French national research agency (ANR project MAQSi).
\end{acknowledgments}


\appendix 

\section{The dipole approximation}
\label{appendixdipole}

If the extension of the qubit wave functions is significantly smaller than the wave length of the phonons involved in the relaxation, we can make the following (dipole) approximation for the phase factor:
\begin{align}
e^{i\vec{q}\cdot(\vec{r}-\vec{r}')} &\approx 1 + i\vec{q}\cdot(\vec{r}-\vec{r}') - \frac{1}{2}[\vec{q}\cdot(\vec{r}-\vec{r}')]^2 \nonumber \\
&\approx 1 + i\vec{q}\cdot\vec{r} - i\vec{q}\cdot\vec{r}'+ (\vec{q}\cdot\vec{r})(\vec{q}\cdot\vec{r}') \nonumber \\
&-\frac{1}{2}[(\vec{q}\cdot\vec{r})^2+(\vec{q}\cdot \vec{r}')^2]\,.
\end{align}
The matrix element in Eq. (\ref{eqgamma3D}) can then be expanded as:
\begin{align}
&\Big|\langle \zero |e^{i\vec{q}\cdot \vec{r}} \Delta H(\epsilon_{\alpha\vec{q}}) | \one \rangle \Big|^2 
\approx {\tilde S}{\tilde S}^* + i \sum_k q_k ({\tilde R}_k {\tilde S}^* - {\tilde R}_k^* {\tilde S}) \nonumber \\
&+ \sum_{k,k'}q_k q_{k'}\left[{\tilde R}_k {\tilde R}_{k'}^*-\frac{1}{2}({\tilde T}_{kk'} {\tilde S}^* + {\tilde T}_{kk'}^* {\tilde S})\right], 
\label{eqM2dipole}
\end{align}
where $q_k$ is the component of $\vec{q}$ on axis $k \in \{x,y,z\}$, and ${\tilde S}, {\tilde R}_k, {\tilde T}_{kk'}$ are defined as:
\begin{subequations}
\begin{align}
{\tilde S}&=\langle \zero |\Delta  H | \one \rangle
= \sum_{i,j} \Delta H_{ij} S_{ij} \\  
{\tilde R}_k&=\langle \zero |\Delta H r_k| \one \rangle
=\sum_{i,j} \Delta H_{ij} R^k_{ij} \\
{\tilde T}_{kk'}&=\langle \zero |\Delta H r_kr_{k'}| \one \rangle
=\sum_{i,j} \Delta H_{ij} T^{kk'}_{ij},
\end{align}
\end{subequations}
with $\Delta H\equiv\Delta H_s(\epsilon_{\alpha\vec{q}})$, and $S_{ij}$, $R^k_{ij}$ and $T^{kk'}_{ij}$ given by Eq. (\ref{eqmoments}).

\section{Convergence of the numerical relaxation rate calculations}
\label{appendixConvergence}

Eq. (\ref{eqGammanum}) is expected to diverge when $\omega$ approaches the edge of a parabolic phonon sub-band with  $v_{\alpha 0}=0$. This is a known pitfall of Fermi Golden rule, whose transition rates are proportional to the phonon density of states, which then behaves as $1/\sqrt{\omega-\omega_{\alpha 0}}$. 

In order to understand the implications of this result, we consider (as in section \ref{sectionbulklimit}) a qubit embedded in a homogeneous nanowire with cross-sectional area $S$. When $S$ increases, the density of phonon sub-bands also increases $\propto S$ [Eq. (\ref{eqNph})], but the effective width of each $\propto 1/(S\sqrt{\omega-\omega_{\alpha 0}})$ peak in the relaxation rate decreases as $1/S^2$. Therefore, $\Gamma_{\rm ph}$ does only converge ``in measure'' to $\Gamma_{\rm ph}^{\rm 3D}$ when $S\to\infty$. In order to smooth out the convergence, we have introduced a cut-off $|q|>2\pi/\lambda_{\rm max}$ in Eq. (\ref{eqGammanum}), where $\lambda_{\rm max}=5.431$ cm is very large with respect to all dimensions of the system. This cut-off was enforced in section \ref{sectionconfinement}.

\begin{figure}
\includegraphics[width=1.0\columnwidth]{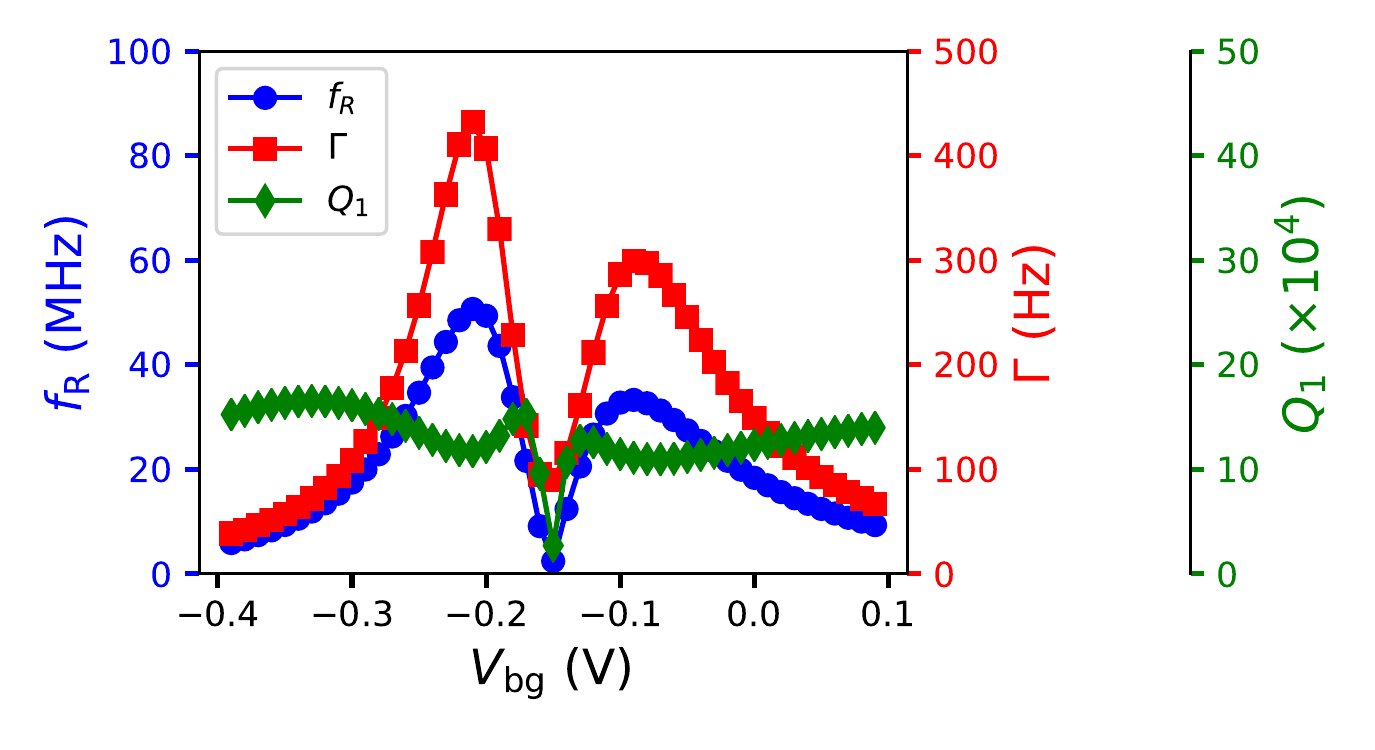}
\caption{Rabi frequency $f_{\rm R}$, total relaxation rate $\Gamma=\Gamma_{\rm ph}^{\rm 3D}+\Gamma_{\rm jn}$ and quality factor $Q_1=f_{\rm R}/\Gamma$ as a function of the back gate voltage $V_{\rm bg}$ ($\theta=45^\circ$, $\varphi=0$, $\omega/(2\pi)=10$ GHz). The Johnson-Nyquist relaxation rate was computed for a resistance $R=200\,\Omega$ and scales $\propto R$.}
\label{figQ1JN}
\end{figure}

From a practical point of view, the van-Hove singularities in the vibrational DoS may also be smoothed by phonon scattering (disorder and phonon-phonon interactions). The mean free path of low-energy acoustic phonons can, however, remain very long at low temperature.\cite{Kargar16}

\section{Figures of merit of the qubit in the presence of Johnson-Nyquist noise}
\label{appendixJN}

\begin{figure}
\includegraphics[width=0.8\columnwidth]{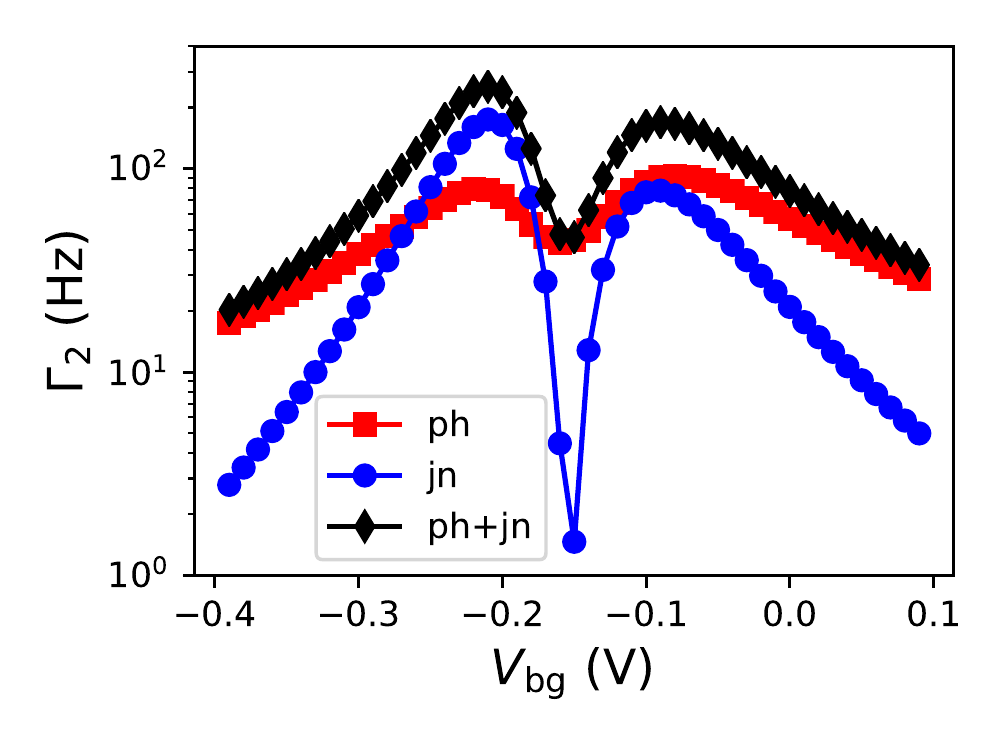}
\caption{Phonon $\Gamma_{2,\rm ph}$, Johnson-Nyquist $\Gamma_{2,\rm jn}$ and total decoherence rate $\Gamma_2=\Gamma_{2,\rm ph}+\Gamma_{2,\rm jn}$ as a function of back gate voltage ($\theta=45^\circ$, $\varphi=0$, $\omega/(2\pi)=10$ GHz).}
\label{figG2JN}
\end{figure}

\begin{figure}
\includegraphics[width=1.0\columnwidth]{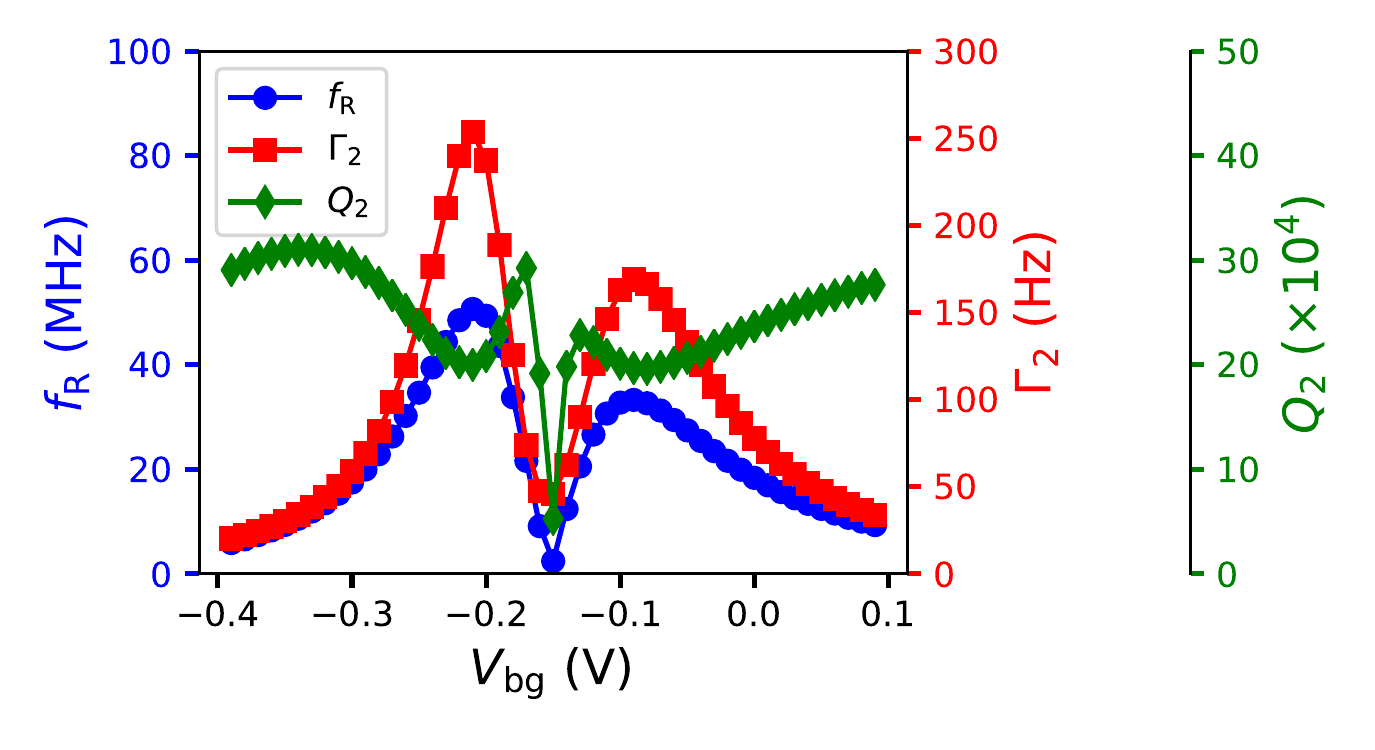}
\caption{Quality factor $Q_2=f_{\rm R}/\Gamma_2$ as a function of back gate voltage ($\theta=45^\circ$, $\varphi=0$, $\omega/(2\pi)=10$ GHz).}
\label{figQ2JN}
\end{figure}

We plot in Fig. \ref{figQ1JN} the Rabi frequency $f_{\rm R}$, the total relaxation rate $\Gamma=\Gamma_{\rm ph}^{\rm 3D}+\Gamma_{\rm jn}$ and the quality factor $Q_1=f_{\rm R}/\Gamma$ as a function of the back gate voltage $V_{\rm bg}$. This figure is the counterpart of Fig. \ref{figoptimal}b including the Johnson-Nyquist contribution computed for a resistance $R=200\,\Omega$. As discussed in the main text, Johnson-Nyquist noise increases the relaxation rate away from  $V_{\rm bg}=-0.15$ V and moves the optimal $Q_1$ farther from that point (because the Rabi frequency is also zero at $V_{\rm bg}=-0.15$ V where the qubit decouples from the electrical noise). 

Phonons and Johnson-Nyquist noise also induce decoherence. The Johnson-Nyquist decoherence rate is $\Gamma_{2,\rm jn}=T_{2,\rm jn}^{-1}=\Gamma_{\rm jn}/2+T_{2,\rm jn}^{*-1}$, where:
\begin{equation}
T_{2,\rm jn}^{*-1}=\frac{2\pi}{\hbar}\frac{R}{R_0}|D_{\one\one}-D_{\zero\zero}|^2k_BT\,,
\end{equation}
with $D_{\zero\zero}=\langle\zero|D_{\rm fg}|\zero\rangle$ and $D_{\one\one}=\langle\one|D_{\rm fg}|\one\rangle$. As for phonons, the decoherence rate is simply $\Gamma_{2,\rm ph}=T_{2,\rm ph}^{-1}=\Gamma_{\rm ph}/2$ because the longitudinal spectral function $S(\omega)$ of phonons is ``super-ohmic'' [$S(\omega=0)=0$] and $T_2^{*-1}\propto S(0)$ in the Bloch-Redfield theory\cite{Loss05,Maier13,Hu14} {\CORRRR (Strictly speaking, $S(0)=0$ for one-phonon processes in 3D, but not necessarily for two-phonon processes,\cite{Kornich14} which may give rise to a finite dephasing time).}

The phonon $\Gamma_{2,\rm ph}$, Johnson-Nyquist $\Gamma_{2,\rm jn}$ and total decoherence rate $\Gamma_2=\Gamma_{2,\rm ph}+\Gamma_{2,\rm jn}$ are plotted as a function of back gate voltage in Fig. \ref{figG2JN}. The quality factor $Q_2=f_{\rm R}/\Gamma_2$ is plotted as a function of $V_{\rm bg}$ in Fig. \ref{figQ2JN}. 

It is clear from Figs. \ref{figQ1JN} and \ref{figQ2JN} that the qubit is limited by relaxation rather than decoherence with the above assumptions. This results from the fact that the spectral densities of both phonons and Johnson-Nyquist noise are small or even zero at low frequency and temperature. 
The situation will be opposite in the presence of a charge noise with a $1/\omega^\alpha$ tail, as decoherence will be much faster.\cite{Paladino14} The modeling of charge noise in such qubits goes, however, far beyond the present work. Hole spin qubits being sensitive to electrical and charge noise owing to the strong spin-orbit coupling in the valence band, this calls for a careful design of devices and electronics around.



\begin{thebibliography}{72}
\bibitem [{not({\natexlab{a}})}]{noteDelta}%
  \BibitemOpen
  \href@noop {} {} \bibinfo {note} {We still neglect here
  the strain dependence of the spin-orbit splitting $\Delta$ between the
  $J=3/2$ and $J=1/2$ multiplets, which is small in Si ($\Delta$ decreases by
  0.75 meV for a $\varepsilon_{XX}=\varepsilon_{YY}=\varepsilon_{ZZ}=1\%$
  hydrostatic strain according to an all-electron density functional theory
  calculation).}\BibitemShut {Stop}%
\bibitem [{not({\natexlab{b}})}]{notedefU}%
  \BibitemOpen
  \href@noop {} {} \bibinfo {note} {$U_{\rm el}$ is here
  defined as: \begin{align} U_{\rm el}&=\frac{1}{2WH}\int_{\rm
  Si}dydz\,c_{11}\left(|\varepsilon_{XX}|^2+|\varepsilon_{YY}|^2+|\varepsilon_{ZZ}|^2\right)
  \nonumber \\
  &+c_{12}\left(\varepsilon_{XX}\varepsilon_{YY}^*+\varepsilon_{XX}\varepsilon_{ZZ}^*+\varepsilon_{YY}\varepsilon_{ZZ}^*+\rm{c.c.}\right)
  \nonumber \\
  &+4c_{44}\left(|\varepsilon_{YZ}|^2+|\varepsilon_{XZ}|^2+|\varepsilon_{XY}|^2\right)\,,
  \nonumber \end{align} where the integral runs over the cross section of the
  silicon wire and the strains in a given phonon sub-band are computed for a
  one phonon excitation.}\BibitemShut {Stop}%
\bibitem [{not({\natexlab{c}})}]{notephonons}%
  \BibitemOpen
  \href@noop {} {} \bibinfo {note} {When $q\to0$, the
  displacements of the three acoustic branches are always homogeneous over the
  whole unit cell, as they are representative of the global translational
  invariance of the system. When increasing $q$, they tend to relocalize in the
  encapsulation material when $\kappa<1$ (phonon depletion in the wire), and in
  silicon when $\kappa>1$ (phonon reconfinement).}\BibitemShut {Stop}%
\bibitem [{not({\natexlab{d}})}]{notephonons2}%
  \BibitemOpen
  \href@noop {} {} \bibinfo {note} {When increasing $L_{\rm
  cell}$, the 3D-like regime $\Gamma_{\rm ph}\propto 1/v_{\rm enc}^5$ lasts up
  to larger $\kappa$ (hence smaller relaxation rates), before turning to the
  1D-like regime $\Gamma_{\rm ph}\propto 1/(v_{\rm enc}^3L_{\rm
  cell}^2)$.}\BibitemShut {Stop}%
\end{thebibliography}

%

\widetext
\clearpage
\begin{center}
\textbf{\large Supplementary material for ``Hole-phonon interactions in quantum dots: Effects of phonon confinement and encapsulation materials on spin-orbit qubits''}
\end{center}
\setcounter{equation}{0}
\setcounter{figure}{0}
\setcounter{table}{0}
\setcounter{section}{0}
\makeatletter
\renewcommand{\thesection}{S\arabic{section}}  
\renewcommand{\theequation}{S\arabic{equation}}  
\renewcommand{\thefigure}{S\arabic{figure}} 
\renewcommand{\thetable}{S\Roman{table}} 
\renewcommand{\bibnumfmt}[1]{[S#1]}
\renewcommand{\citenumfont}[1]{S#1}
\renewcommand\appendixname{Supplementary}

In this supplementary material, we discuss the relaxation rates for strongly confined 2D and 1D phonon band structures. We choose the cubic axes $\vec{x}\parallel[100]$, $\vec{y}\parallel[010]$ and $\vec{z}\parallel[001]$ as the reference frame. The azimuthal angle $\theta$ is measured with respect to $\vec{z}$, and the polar angle $\varphi$ with respect to $\vec{x}$. 

In the sums over coordinates, the $x$, $y$ and $z$ axes are labeled by an integer ($1$ stands for $x$, $2$ for $y$ and $3$ for $z$).

\section{Relaxation rate in thin films}
\label{section_film}

We consider a thin film with thickness $L$, normal to $\hat{\vec{n}}(\theta_0,\varphi_0)$. We define the two in-plane unit vectors $\hat{\vec{q}}_1(\theta_0+\pi/2,\varphi_0)$, and $\hat{\vec{q}}_2(\pi/2,\varphi_0+\pi/2)$. We assume Born-von-Karman periodic boundary conditions at the surface of the film. The resulting phonon band structure is then the bulk band structure sampled at wave vectors:
\begin{equation}
\vec{q}=\frac{2\pi n}{L}\hat{\vec{n}}+q_{\parallel}(\cos\gamma\hat{\vec{q}}_1+\sin\gamma\hat{\vec{q}}_2)\,,
\end{equation}
where $n$ is an integer and $\gamma$ is the angle between the in-plane wave vector $\vec{q}_{\parallel}$ and $\hat{\vec{q}}_1$. Each value of $n$ defines three phonon sub-bands, sampled in the bulk LA, TA1 and TA2 branches. The acoustic branches of the film ($\omega_{\alpha\vec{q}_\parallel}\to 0$ when $\vec{q}_\parallel\to\vec{0}$) are the $n=0$ sub-bands. In this approximation, the wave function of the acoustic branches are, therefore, homogeneous in the thickness of the film. The other sub-bands tend to a finite $\omega_{\alpha\vec{0}}$. For strong enough confinement, only the $n=0$ sub-bands can hence couple to the qubit. The polarization vectors of the LA and TA branches of the film are:
\begin{align}
\hat{c}_{l,x}&= \cos \gamma \cos\theta_0 \cos\varphi_0 - \sin \gamma \sin \varphi_0 \nonumber \\
\hat{c}_{l,y}&= \cos \gamma \cos\theta_0 \sin\varphi_0 + \sin \gamma \cos \varphi_0 \nonumber \\
\hat{c}_{l,z}&= -\cos \gamma \sin\theta_0 \nonumber \\
\hat{c}_{t1,x}&= \sin\theta_0 \cos\varphi_0 \nonumber \\
\hat{c}_{t1,y}&= \sin\theta_0 \sin\varphi_0 \nonumber \\
\hat{c}_{t1,z}&= \cos\theta_0 \nonumber \\
\hat{c}_{t2,x}&= \cos \gamma \sin\varphi_0 + \sin \gamma \cos \varphi_0 \cos \theta_0 \nonumber \\
\hat{c}_{t2,y}&= -\cos \gamma \cos \varphi_0 + \sin \gamma \sin \varphi_0 \cos \theta_0 \nonumber \\
\hat{c}_{t2,z}&= -\sin \gamma \sin \theta_0\,.
\end{align}
The integration can then be completed in Eq. (8) of the main text, the lines $\omega_{\alpha\vec{q}_\parallel}=\omega$ being circles. We end up with Eq. (25) for $\Gamma^{\rm 2D}_{\rm ph}$. The value of the $A_n$'s, $B_n$'s and $\Lambda$ parameters are given below for $(001)$, $(110)$ and $(111)$ films.

\subsection{$(001)$ films}

For a $(001)$ film, $\theta_0=0$ and $\varphi_0=0$. With indices $i,j,k,l,m\in\{1,2\}$, the 11 $A_n$ terms are:
\begin{equation}
A_1=O^{11}_{1111}
   +O^{22}_{2222}
\end{equation}
\begin{equation}	
A_2=O^{11}_{2222}
   +O^{22}_{1111}
\end{equation}
\begin{equation}		
A_3=O^{11}_{3333}
   +O^{22}_{3333}
\end{equation}
\begin{equation}		
A_4=\sum_{i \neq j} 
  O^{11}_{iijj}
 +O^{22}_{iijj}
\end{equation}
\begin{equation}		
A_5=\sum_{i} 
  O^{ii}_{ii33}
 +O^{ii}_{33ii}
\end{equation}
\begin{equation}		
A_6=\sum_{i\neq j} 
  O^{ii}_{jj33}
 +O^{ii}_{33jj}
\end{equation}
\begin{equation}		
A_7=\sum_{i, j\neq k, l \neq m} 
  O^{ii}_{jklm}
\end{equation}
\begin{equation}		
A_8=\sum_{i \neq j,k \neq l,m} O^{ij}_{mmkl} + O^{ji}_{klmm}
\end{equation}
\begin{equation}		
A_9=\sum_{i \neq j,k \neq l} O^{ij}_{33kl} + O^{ji}_{kl33}
\end{equation}
\begin{equation}		
A_{10}=\sum_{i}
  O^{ii}_{i3i3}
+ O^{ii}_{i33i}
+ O^{ii}_{3ii3}
+ O^{ii}_{3i3i}
\end{equation}
\begin{equation}		
A_{11}=\sum_{i \neq j, k \neq l} 
  O^{ii}_{j3j3}
+ O^{ii}_{j33j}
+ O^{ii}_{3jj3}
+ O^{ii}_{3j3j}
+ O^{ij}_{k3l3} + O^{ij}_{k33l}
+ O^{ij}_{3kl3} + O^{ij}_{3k3l}\,.
\end{equation}
The parameters for longitudinal acoustic phonons are:
$\Lambda^{\rm A}_{1l}=\frac{8 a^2 + 20 ab + 17 b^2}{8}$;
$\Lambda^{\rm A}_{2l}= \frac{8 a^2 - 4 ab + 5 b^2}{8}$;
$\Lambda^{\rm A}_{3l}= a^2 - 2 ab + b^2$;                
$\Lambda^{\rm A}_{4l}= \frac{8a^2 + 8ab - 7 b^2}{8}$;    
$\Lambda^{\rm A}_{5l}= \frac{4a^2 + ab - 5 b^2}{4}$;    
$\Lambda^{\rm A}_{6l}= \frac{4a^2 - 5ab + b^2}{4}$;   
$\Lambda^{\rm A}_{7l}= \frac{3 d^2}{8}$;                
$\Lambda^{\rm A}_{8l}= \sqrt{3}d\frac{2a +b}{8}$;        
$\Lambda^{\rm A}_{9l}= \sqrt{3}d\frac{a-b}{4}$;          
$\Lambda^{\rm A}_{10l}= 0$;         
$\Lambda^{\rm A}_{11l}= 0$. \\
The parameters for transverse acoustic phonons are:
$\Lambda^{\rm A}_{1t}=\frac{9 b^2}{8}$;
$\Lambda^{\rm A}_{2t}=\frac{9 b^2}{8}$;
$\Lambda^{\rm A}_{3t}=0$;                             
$\Lambda^{\rm A}_{4t}=\frac{-9 b^2}{8}$; 
$\Lambda^{\rm A}_{5t}=0$;             
$\Lambda^{\rm A}_{6t}=0$;             
$\Lambda^{\rm A}_{7t}=\frac{3 d^2}{8}$; 
$\Lambda^{\rm A}_{8t}=0$;             
$\Lambda^{\rm A}_{9t}=0$;               
$\Lambda^{\rm A}_{10t}=\frac{9 d^2}{16}$;
$\Lambda^{\rm A}_{11t}=\frac{3 d^2}{16}$. \\
There are 6 $B_n$ terms:
\begin{equation}		
B_1=S_{11}S^*_{11} + S_{22}S^*_{22}
\end{equation}
\begin{equation}	
B_2= S_{33}S^*_{33}
\end{equation}
\begin{equation}	
B_3= S_{11}S^*_{22} + S_{22}S^*_{11}
\end{equation}
\begin{equation}	
B_4= S_{11}S^*_{33} + S_{33}S^*_{11}
+    S_{22}S^*_{33} + S_{33}S^*_{22}
\end{equation}
\begin{equation}	
B_5= S_{12}S^*_{12} + S_{12}S^*_{21}
+    S_{21}S^*_{12} + S_{21}S^*_{21}
\end{equation}
\begin{equation}	
B_6= \sum_{i}S_{i3}S^*_{i3} + S_{i3}S^*_{3i}
+            S_{3i}S^*_{i3} + S_{i3}S^*_{3i} 
\end{equation}
The parameters for longitudinal acoustic phonons are:
$\Lambda^{\rm B}_{1l}=\frac{8a^2 + 8ab + 11b^2}{4}$;
$\Lambda^{\rm B}_{2l}=2(a-b)^2$;                    
$\Lambda^{\rm B}_{3l}= \frac{8a^2 + 8ab - 7b^2}{4}$;
$\Lambda^{\rm B}_{4l}=2a^2-ab-b^2$;                
$\Lambda^{\rm B}_{5l}=\frac{3 d^2}{4}$;           
$\Lambda^{\rm B}_{6l}=0$. \\                   
The parameters for transverse acoustic phonons are:
$\Lambda^{\rm B}_{1t}=\frac{9b^2}{4}$;           
$\Lambda^{\rm B}_{2t}=0$;                         
$\Lambda^{\rm B}_{3t}=\frac{-9b^2}{4}$;     
$\Lambda^{\rm B}_{4t}=0$;                     
$\Lambda^{\rm B}_{5t}=\frac{3 d^2}{4}$;       
$\Lambda^{\rm B}_{6t}=\frac{3 d^2}{4}$.         

\subsection{$(110)$ films}

For a $(110)$ film, $\theta_0=\pi/2$ and $\varphi_0=\pi/4$. With indices $i,j,k,l,m\in\{1,2\}$, the 21 $A_n$ terms are:
\begin{equation}
A_1=\sum_{i,j,k} (-1)^{(i+j)} O^{ij}_{kkkk}
\end{equation}
\begin{equation}
A_2=\sum_{i,j} (-1)^{(i+j)} O^{ij}_{3333}
\end{equation}    
\begin{equation}
A_3=\sum_{i,j,k \neq l} (-1)^{(i+j)} O^{ij}_{kkll}
\end{equation}
\begin{equation}
A_4=\sum_{i,j,k} (-1)^{(i+j)} (O^{ij}_{kk33} + O^{ij}_{33kk})                 
\end{equation} 
\begin{equation}
A_5=\sum_{i,j,k,l\neq m} (-1)^{(i+j)}(O^{ij}_{kklm} + O^{ij}_{lmkk})
\end{equation}
\begin{equation}
A_6=\sum_{i,j,k\neq l} (-1)^{(i+j)} (O^{ij}_{33kl} + O^{ij}_{kl33})
\end{equation}
\begin{equation}
A_7=\sum_{i,j,k} 
 (-1)^{(i+j)} (O^{i3}_{jjk3} + O^{i3}_{jj3k}
+              O^{i3}_{k3jj} + O^{i3}_{3kjj}
+              O^{3i}_{jjk3} + O^{3i}_{jj3k}
+              O^{3i}_{k3jj} + O^{3i}_{3kjj}) 
\end{equation}
\begin{equation}
A_8=\sum_{k,l\neq m} 
  O^{33}_{kklm} 
+ O^{33}_{lmkk}
\end{equation}
\begin{equation}
A_9=\sum_{i,j} (-1)^{(i+j)}
(O^{i3}_{33j3} + O^{i3}_{333j} + O^{i3}_{j333} + O^{i3}_{3j33} 
+O^{3i}_{33j3} + O^{3i}_{333j} + O^{3i}_{j333} + O^{3i}_{3j33})
-\sum_{i\neq j} O^{33}_{33ij} + O^{33}_{ij33} 
\end{equation}
\begin{equation}
A_{10}=\sum_{i,j,k \neq l} (-1)^{(i+j)}(O^{ij}_{klkl} + O^{ij}_{kllk})
\end{equation}
\begin{equation}
A_{11}=\sum_{i,j,k \neq l} (-1)^{(i+j+1)}(O^{ij}_{k3l3} + O^{ij}_{k33l} + O^{ij}_{3kl3} + O^{ij}_{3k3l})
\end{equation}
\begin{equation}
A_{12}=\sum_{i,j,k} (-1)^{(i+j)}(O^{ij}_{k3k3}+O^{ij}_{k33k}+O^{ij}_{3kk3}+O^{ij}_{3k3k})
\end{equation}
\begin{equation}
A_{13}=\sum_{i} O^{33}_{iiii}
\end{equation}
\begin{equation}
A_{14}=O^{33}_{3333}
\end{equation}
\begin{equation}
A_{15}=\sum_{i \neq j} O^{33}_{iijj}
\end{equation}
\begin{equation}
A_{16}=\sum_{i} O^{33}_{ii33} 
             +  O^{33}_{33ii}
\end{equation}
\begin{equation}
A_{17}=\sum_{i} O^{33}_{i3i3}
+               O^{33}_{i33i}
+               O^{33}_{3ii3}
+               O^{33}_{3i3i} 
\end{equation}
\begin{equation}
A_{18}=\sum_{i \neq j} O^{33}_{ijij} + O^{33}_{ijji}
\end{equation}
\begin{equation}
A_{19}=\sum_{i \neq j} O^{33}_{i3j3} + O^{33}_{i33j} + O^{33}_{3ij3} + O^{33}_{3i3j} 
\end{equation}
\begin{equation}
A_{20}=\sum_{i, j\neq k, l} (-1)^{(i+l+1)} (
 O^{i3}_{jkl3} + O^{i3}_{jk3l}
+O^{i3}_{l3jk} + O^{i3}_{3ljk}
+O^{3i}_{jkl3} + O^{3i}_{jk3l}
+O^{3i}_{l3jk} + O^{3i}_{3ljk}) 
\end{equation}
\begin{equation}
A_{21}=\sum_{i \neq j} O^{33}_{ij33}
+                      O^{33}_{33ij}\,.
\end{equation}
The parameters for longitudinal acoustic phonons are:
$\Lambda^{\rm A}_{1l}=\Lambda^{\rm A}_{3l}=\frac{32 a^2 + 8 ab + 5 b^2}{64}$; 
$\Lambda^{\rm A}_{2l}= \frac{8 a^2 - 4 ab + 5 b^2}{16}$;                  
$\Lambda^{\rm A}_{4l}= \frac{16a^2 - 2ab - 5 b^2}{32}$;                   
$\Lambda^{\rm A}_{5l}= -\sqrt{3}d\frac{12a +3b}{64}$;                     
$\Lambda^{\rm A}_{6l}= \sqrt{3}d\frac{3b-6a}{32}$;                     
$\Lambda^{\rm A}_{7l}= \sqrt{3}d\frac{4a-b}{32}$;                      
$\Lambda^{\rm A}_{8l}= \sqrt{3}d\frac{b-4a}{32}$;                      
$\Lambda^{\rm A}_{9l}= \sqrt{3}d\frac{2a+b}{16}$;                         
$\Lambda^{\rm A}_{10l}= \frac{15d^2}{64}$;                                
$\Lambda^{\rm A}_{11l}=\Lambda^{\rm A}_{12l}=\Lambda^{\rm A}_{20l} = \frac{3d^2}{32}$;          
$\Lambda^{\rm A}_{13l}=\Lambda^{\rm A}_{15l} = \frac{32a^2 - 40ab + 17 b^2}{32}$;   
$\Lambda^{\rm A}_{14l}=\frac{8a^2 + 20ab + 17 b^2}{8}$;                      
$\Lambda^{\rm A}_{16l}=\frac{16a^2 + 10ab - 17 b^2}{16}$;                  
$\Lambda^{\rm A}_{17l}=\Lambda^{\rm A}_{18l}= \frac{3d^2}{16}$;                       
$\Lambda^{\rm A}_{19l}= -\frac{3d^2}{16}$;                                 
$\Lambda^{\rm A}_{21l}= 0$. \\                                                   
The parameters for transverse acoustic phonons are:
$\Lambda^{\rm A}_{1t}=\frac{63 b^2}{64}$;                      
$\Lambda^{\rm A}_{2t}=\frac{9 b^2}{16}$;                       
$\Lambda^{\rm A}_{3t}=\frac{-45 b^2}{64}$;                                    
$\Lambda^{\rm A}_{4t}=\frac{-9 b^2}{32}$;                      
$\Lambda^{\rm A}_{5t}=-\sqrt{3}d\frac{3b}{64}$;                
$\Lambda^{\rm A}_{6t}=\Lambda^{\rm A}_{7t}=\sqrt{3}d\frac{3b}{32}$;     
$\Lambda^{\rm A}_{8t}=\Lambda^{\rm A}_{9t}=0$;                          
$\Lambda^{\rm A}_{10t}=\Lambda^{\rm A}_{11t}=\frac{3 d^2}{64}$;         
$\Lambda^{\rm A}_{12t}=\frac{9 d^2}{64}$;                      
$\Lambda^{\rm A}_{13t}=\frac{27 b^2}{32}$;                    
$\Lambda^{\rm A}_{14t}=\frac{9 b^2}{8}$;                       
$\Lambda^{\rm A}_{15t}=\frac{-9 b^2}{32}$;           
$\Lambda^{\rm A}_{16t}=\frac{-9 b^2}{16}$;                    
$\Lambda^{\rm A}_{17t}=\frac{15 d^2}{32}$;                    
$\Lambda^{\rm A}_{18t}=\Lambda^{\rm A}_{19t}=\frac{3 d^2}{32}$;        
$\Lambda^{\rm A}_{20t}=0$;                                  
$\Lambda^{\rm A}_{21t}=\sqrt{3}d\frac{3b}{16}$. \\
There are 9 $B_n$ terms:
\begin{equation}		
B_1=S_{11}S^*_{11} + S_{22}S^*_{22}
\end{equation}
\begin{equation}	
B_2=S_{33}S^*_{33}
\end{equation}
\begin{equation}	
B_3=S_{11}S^*_{22} + S_{22}S^*_{11}
\end{equation}
\begin{equation}	
B_4=S_{11}S^*_{33} + S_{33}S^*_{11}
+   S_{22}S^*_{33} + S_{33}S^*_{22} 
\end{equation}
\begin{equation}	
B_5=\sum_{i}S_{ii}S^*_{12} + S_{ii}S^*_{21}
+           S_{12}S^*_{ii} + S_{21}S^*_{ii} 
\end{equation}
\begin{equation}	
B_6=S_{33}S^*_{12} + S_{33}S^*_{21}
+   S_{12}S^*_{33} + S_{21}S^*_{33} 
\end{equation}
\begin{equation}	
B_7=S_{12}S^*_{12} + S_{12}S^*_{21}
+   S_{21}S^*_{12} + S_{21}S^*_{21} 
\end{equation}
\begin{equation}	
B_8=\sum_{i}S_{i3}S^*_{i3} + S_{i3}S^*_{3i}
+           S_{3i}S^*_{i3} + S_{3i}S^*_{3i} 
\end{equation}
\begin{equation}	
B_9=\sum_{i\neq j}S_{i3}S^*_{j3} + S_{i3}S^*_{3j}
+                 S_{3i}S^*_{j3} + S_{3i}S^*_{3j} 
\end{equation}
The parameters for longitudinal acoustic phonons are:
$\Lambda^{\rm B}_{1l}=\Lambda^{\rm B}_{3l}=\frac{32a^2 - 16ab + 11b^2}{16}$; 
$\Lambda^{\rm B}_{2l}=\frac{ 8a^2 +  8ab + 11b^2}{ 4}$;           
$\Lambda^{\rm B}_{4l}=\frac{16a^2 + 4ab - 11b^2}{8}$;          
$\Lambda^{\rm B}_{5l}=-\sqrt{3}d\frac{8a+b}{16}$;               
$\Lambda^{\rm B}_{6l}=\sqrt{3}d\frac{b-4a}{8}$;                
$\Lambda^{\rm B}_{7l}= \frac{9d^2}{16}$;                          
$\Lambda^{\rm B}_{8l}= \frac{3d^2}{ 8}$;                           
$\Lambda^{\rm B}_{9l}=-\frac{3d^2}{ 8}$. \\                           
The parameters for transverse acoustic phonons are:
$\Lambda^{\rm B}_{1t}=\frac{45b^2}{16}$;             
$\Lambda^{\rm B}_{2t}=\frac{ 9b^2}{ 4}$;          
$\Lambda^{\rm B}_{3t}=\frac{-27b^2}{16}$;    
$\Lambda^{\rm B}_{4t}=\frac{-9b^2}{8}$;             
$\Lambda^{\rm B}_{5t}=-\sqrt{3}d \frac{3 b}{16}$;   
$\Lambda^{\rm B}_{6t}=\sqrt{3}d \frac{3 b}{8}$;     
$\Lambda^{\rm B}_{7t}=\frac{3d^2}{16}$;               
$\Lambda^{\rm B}_{8t}=\frac{3d^2}{ 4}$;              
$\Lambda^{\rm B}_{9t}=0$.                     

\subsection{$(111)$ films}

For a $(111)$ film, $\theta_0=\arccos(\sqrt{3}/3)$ and $\varphi_0=\pi/4$. With indices $i,j,k,l,m\in\{1,2,3\}$, the 25 $A_n$ terms are: 
\begin{equation}
A_1=\sum_{i} O^{ii}_{iiii}
\end{equation}
\begin{equation}	
A_2=\sum_{i\neq j} O^{ii}_{jjjj}
\end{equation}
\begin{equation}		
A_3=\sum_{i \neq j} O^{ii}_{iijj} + O^{ii}_{jjii}
\end{equation}
\begin{equation}		
A_4=\sum_{i \neq j \neq k} O^{ii}_{jjkk}
\end{equation}
\begin{equation}		
A_5=\sum_{i \neq j} O^{ii}_{iiij} + O^{ii}_{iiji} + O^{ii}_{ijii} + O^{ii}_{jiii}
\end{equation}
\begin{equation}		
A_6=\sum_{i \neq j \neq k} O^{ii}_{iijk} + O^{ii}_{jkii}
\end{equation}
\begin{equation}		
A_7=\sum_{i \neq j} O^{ii}_{jjij} + O^{ii}_{jjji} + O^{ii}_{ijjj} + O^{ii}_{jijj} 
\end{equation}
\begin{equation}		
A_8=\sum_{i \neq j \neq k} O^{ii}_{jjik} + O^{ii}_{jjki} + O^{ii}_{ikjj} + O^{ii}_{kijj} 
\end{equation}
\begin{equation}		
A_9=\sum_{i \neq j \neq k} O^{ii}_{jjjk} + O^{ii}_{jjkj} + O^{ii}_{jkjj} + O^{ii}_{kjjj} 
\end{equation}
\begin{equation}		
A_{10}=\sum_{i \neq j} O^{ii}_{ijij} + O^{ii}_{ijji} + O^{ii}_{jiij} + O^{ii}_{jiji} 
\end{equation}
\begin{equation}		
A_{11}=\sum_{i \neq j \neq k} O^{ii}_{jkjk} + O^{ii}_{jkkj}  
\end{equation}
\begin{equation}		
A_{12}=\sum_{i \neq j \neq k} O^{ii}_{ijik} + O^{ii}_{ijki} + O^{ii}_{jiik} + O^{ii}_{jiki} 
\end{equation}
\begin{equation}		
A_{13}=\sum_{i \neq j \neq k} O^{ii}_{ijjk} + O^{ii}_{ijkj} + O^{ii}_{jijk} + O^{ii}_{jikj} + O^{ii}_{jkij} + O^{ii}_{jkji} + O^{ii}_{kjij} + O^{ii}_{kjji} 
\end{equation}
\begin{equation}		
A_{14}=\sum_{i \neq j} O^{ij}_{iiii} + O^{ji}_{iiii}
\end{equation}
\begin{equation}		
A_{15}=\sum_{i \neq j \neq k} O^{ij}_{kkkk}
\end{equation}
\begin{equation}		
A_{16}=\sum_{i \neq j} O^{ij}_{iijj} + O^{ji}_{iijj} 
\end{equation}
\begin{equation}		
A_{17}=\sum_{i \neq j \neq k} O^{ij}_{iikk} + O^{ji}_{kkii} + O^{ij}_{kkii} + O^{ji}_{iikk}
\end{equation}
\begin{equation}		
A_{18}=\sum_{i \neq j}
  O^{ij}_{iiij} + O^{ij}_{iiji}
+ O^{ij}_{ijii} + O^{ij}_{jiii}
+ O^{ji}_{iiij} + O^{ji}_{iiji} 
+ O^{ji}_{ijii} + O^{ji}_{jiii} 
\end{equation}
\begin{equation}		
A_{19}=\sum_{i \neq j \neq k}
  O^{ij}_{iiik} + O^{ij}_{iiki}
+ O^{ij}_{ikii} + O^{ij}_{kiii}
+ O^{ji}_{iiik} + O^{ji}_{iiki}    
+ O^{ji}_{ikii} + O^{ji}_{kiii}
\end{equation}
\begin{equation}		
A_{20}=\sum_{i \neq j \neq k, l \neq m, m \neq k}
  O^{ij}_{llmk} + O^{ij}_{llkm}
+ O^{ij}_{mkll} + O^{ij}_{kmll}
\end{equation}
\begin{equation}		
A_{21}=\sum_{i \neq j}
  O^{ij}_{ijij} + O^{ij}_{ijji}
+ O^{ij}_{jiij} + O^{ij}_{jiji}
\end{equation}
\begin{equation}		
A_{22}=\sum_{i \neq j \neq k}
  O^{ij}_{ikik} + O^{ij}_{ikki}
+ O^{ij}_{kiik} + O^{ij}_{kiki}
+ O^{ji}_{ikik} + O^{ji}_{ikki}  
+ O^{ji}_{kiik} + O^{ji}_{kiki} 
\end{equation}
\begin{align}		
A_{23}=\sum_{i \neq j \neq k}
&O^{ij}_{ijik} + O^{ij}_{ijki} + O^{ij}_{jiik}
+O^{ij}_{jiki} + O^{ij}_{ikij} + O^{ij}_{kiij}
+O^{ij}_{ikji} + O^{ij}_{kiji}\nonumber \\
&O^{ji}_{ijik} + O^{ji}_{ijki} + O^{ji}_{jiik}
+O^{ji}_{jiki} + O^{ji}_{ikij} + O^{ji}_{kiij}
+O^{ji}_{ikji} + O^{ji}_{kiji}
\end{align}
\begin{equation}		
A_{24}=\sum_{i \neq j \neq k}
  O^{ij}_{ikjk} + O^{ij}_{ikkj}
+ O^{ij}_{kijk} + O^{ij}_{kikj}
+ O^{ji}_{ikjk} + O^{ji}_{ikkj}
+ O^{ji}_{kijk} + O^{ji}_{kikj}
\end{equation}
\begin{equation}		
A_{25}=\sum_{i \neq j \neq k}
  O^{ij}_{kkij} + O^{ij}_{kkji}
+ O^{ij}_{ijkk} + O^{ij}_{jikk}
\end{equation}
The parameters for longitudinal acoustic phonons are:
$\Lambda^{\rm A}_{1l}= \frac{2 a^2 + 2 ab + b^2}{3}$;   
$\Lambda^{\rm A}_{2l}= \frac{2 a^2 -   ab + b^2}{3}$;     
$\Lambda^{\rm A}_{3l}= \frac{4 a^2 +   ab - b^2}{6}$;     
$\Lambda^{\rm A}_{4l}= \frac{4 a^2 - 2 ab - b^2}{6}$;    
$\Lambda^{\rm A}_{5l}=-\sqrt{3}d\frac{3a + 2b}{18}$;      
$\Lambda^{\rm A}_{6l}= \sqrt{3}d\frac{      b}{18}$;     
$\Lambda^{\rm A}_{7l}=-\sqrt{3}d\frac{6a +  b}{36}$;    
$\Lambda^{\rm A}_{8l}=-\sqrt{3}d\frac{6a - 5b}{36}$;      
$\Lambda^{\rm A}_{9l}= \sqrt{3}d\frac{   -  b}{36}$;      
$\Lambda^{\rm A}_{10l}=              \frac{2d^2}{ 9}$;    
$\Lambda^{\rm A}_{11l}=\Lambda^{\rm A}_{12l}= \frac{ d^2}{18}$;       
$\Lambda^{\rm A}_{13l}=             -\frac{ d^2}{36}$;       
$\Lambda^{\rm A}_{14l}=-\frac{2 a^2 + 2 ab + b^2}{ 6}$;       
$\Lambda^{\rm A}_{15l}=-\frac{2 a^2 - 4 ab + b^2}{ 6}$;   
$\Lambda^{\rm A}_{16l}=-\frac{4 a^2 + 4 ab - b^2}{12}$;    
$\Lambda^{\rm A}_{17l}=-\frac{4 a^2 - 2 ab - b^2}{12}$;    
$\Lambda^{\rm A}_{18l}= \sqrt{3}d\frac{3a + b}{18}$;      
$\Lambda^{\rm A}_{19l}= \sqrt{3}d\frac{     b}{18}$;       
$\Lambda^{\rm A}_{20l}=-\sqrt{3}d\frac{     b}{36}$;       
$\Lambda^{\rm A}_{21l}=              -\frac{7d^2}{36}$;    
$\Lambda^{\rm A}_{22l}=\Lambda^{\rm A}_{22l}= -\frac{ d^2}{36}$;    
$\Lambda^{\rm A}_{24l}=               -\frac{d^2}{18}$;   
$\Lambda^{\rm A}_{25l}= \sqrt{3}d\frac{3a - 2b}{18}$. \\      
The parameters for transverse acoustic phonons are:
$\Lambda^{\rm A}_{1t}= \frac{4 b^2}{3}$;                   
$\Lambda^{\rm A}_{2t}= \frac{5 b^2}{6}$;                   
$\Lambda^{\rm A}_{3t}=-\frac{2 b^2}{3}$;                               
$\Lambda^{\rm A}_{4t}= \frac{- b^2}{6}$;                  
$\Lambda^{\rm A}_{5t}= \Lambda^{\rm A}_{8t}=\Lambda^{\rm A}_{9t}=\sqrt{3}d\frac{b}{36}$;            
$\Lambda^{\rm A}_{6t}=\Lambda^{\rm A}_{7t}=-\sqrt{3}d\frac{b}{18}$; 
$\Lambda^{\rm A}_{10t}=\frac{11 d^2}{72}$;                 
$\Lambda^{\rm A}_{11t}=\frac{7 d^2}{36}$;                 
$\Lambda^{\rm A}_{12t}=-\frac{d^2}{18}$;                    
$\Lambda^{\rm A}_{13t}=-\frac{7 d^2}{72}$;                  
$\Lambda^{\rm A}_{14t}=-\frac{2 b^2}{3}$;                    
$\Lambda^{\rm A}_{15t}=-\frac{ b^2}{6}$;                   
$\Lambda^{\rm A}_{16t}= \frac{7b^2}{12}$;                 
$\Lambda^{\rm A}_{17t}= \frac{ b^2}{12}$;                   
$\Lambda^{\rm A}_{18t}=\Lambda^{\rm A}_{20t}=\sqrt{3}d\frac{b}{36}$; 
$\Lambda^{\rm A}_{19t}=-\sqrt{3}d\frac{b}{18}$;             
$\Lambda^{\rm A}_{21t}=-\frac{d^2}{18}$;              
$\Lambda^{\rm A}_{22t}=-\frac{7d^2}{72}$;               
$\Lambda^{\rm A}_{23t}=\frac{d^2}{36}$;                
$\Lambda^{\rm A}_{24t}=\frac{5d^2}{72}$;                 
$\Lambda^{\rm A}_{25t}=-\sqrt{3}d\frac{b}{18}$. \\      
There are 6 $B_n$ terms:
\begin{equation}		
B_1=\sum_{i} S_{ii}S^*_{ii}
\end{equation}
\begin{equation}	
B_2=\sum_{i \neq j} S_{ii}S^*_{jj}
\end{equation}
\begin{equation}	
B_3=\sum_{i \neq j} 
 S_{ii}S^*_{ij} + S_{ii}S^*_{ji}
+S_{ij}S^*_{ii} + S_{ji}S^*_{ii} 
\end{equation}
\begin{equation}	
B_4=\sum_{i \neq j \neq k} 
 S_{ii}S^*_{jk} + S_{jk}S^*_{ii}
\end{equation}
\begin{equation}	
B_5=\sum_{i \neq j} 
 S_{ij}S^*_{ij} + S_{ij}S^*_{ji}
\end{equation}
\begin{equation}	
B_6=\sum_{i \neq j \neq k} 
 S_{ij}S^*_{ik} + S_{ij}S^*_{ki}
+S_{ji}S^*_{ik} + S_{ji}S^*_{ki} 
\end{equation}
The parameters for longitudinal acoustic phonons are:
$\Lambda^{\rm B}_{1l}=2a^2 + b^2$;               
$\Lambda^{\rm B}_{2l}=\frac{4a^2-b^2}{2}$;        
$\Lambda^{\rm B}_{3l}=-\sqrt{3}d \frac{2a+b}{6}$;
$\Lambda^{\rm B}_{4l}=\sqrt{3}d \frac{b-a}{3}$;  
$\Lambda^{\rm B}_{5l}=\frac{d^2}{2}$;           
$\Lambda^{\rm B}_{6l}=0$. \\                       
The parameters for transverse acoustic phonons are:
$\Lambda^{\rm B}_{1t}=       3b^2$;            
$\Lambda^{\rm B}_{2t}=\frac{-3b^2}{2}$;          
$\Lambda^{\rm B}_{3t}=\Lambda^{\rm B}_{4t}=0$;          
$\Lambda^{\rm B}_{5t}=\frac{  d^2}{2}$;       
$\Lambda^{\rm B}_{6t}=\frac{- d^2}{4}$.       

\section{Relaxation rate in nanowires}
\label{section_wire}

We consider a wire oriented along $\hat{\vec{u}}(\theta_0,\varphi_0)$, with width $l_1$ along $\hat{\vec{q}}_1$ and height $l_2$ along $\hat{\vec{q}}_2$ ($\hat{\vec{q}}_1\perp\hat{\vec{u}}$, $\hat{\vec{q}}_2\perp\hat{\vec{u}}$). We again assume Born-von-Karman periodic boundary conditions at the surface of the wire. The resulting phonon band structure is then the bulk band structure sampled at wave vectors:
\begin{equation}
\vec{q} = q_\parallel \hat{\vec{u}} + \frac{2 \pi n_1}{l_1} \hat{\vec{q}}_1 + \frac{2 \pi n_2}{l_2} \hat{\vec{q}}_2\,,
\end{equation}
where $n_1$ and $n_2$ are integers. Each pair $(n_1, n_2)$ defines three phonon sub-bands, sampled in the bulk LA, TA1 and TA2 branches. The acoustic branches of the film ($\omega_{\alpha\vec{q}_\parallel}\to 0$ when $\vec{q}_\parallel\to\vec{0}$) are the $n_1=n_2=0$ sub-bands. For strong enough confinement, only the $n_1=n_2=0$ sub-bands can couple to the qubit. The integration can then be completed in Eq. (8) of the main text. We end up with Eq. (26) for $\Gamma^{\rm 1D}_{\rm ph}$. The value of the $A_n$'s, $B_n$'s and $\Lambda$ parameters are given below for $[001]$, $[110]$ and $[111]$-oriented wires. Note that the shape of the nanowire is not relevant as the wave function of the acoustic branches are homogeneous in the cross-section of the wire in this approximation. The relaxation rate only depends on the area of this cross-section.

\subsection{$[001]$ wires}

For a $[001]$ wire, $\theta_0=0$ and $\varphi_0=0$. With indices $i,j\in\{1,2\}$, the 4 $A_n$ terms are:
\begin{equation}		
A_1=\sum_{ij} O^{33}_{iijj}
\end{equation}
\begin{equation}	
A_2=O^{33}_{3333}
\end{equation}
\begin{equation}	
A_3=\sum_{i} O^{33}_{ii33} + O^{33}_{33ii}
\end{equation}
\begin{equation}	
A_4=\sum_{i} O^{33}_{i3i3} + O^{33}_{i33i} + O^{33}_{3ii3} + O^{33}_{3i3i}\,.
\end{equation}
The parameters for longitudinal acoustic phonons are:
$\Lambda^{\rm A}_{1l}=2a^2 - 4ab + 2b^2$;
$\Lambda^{\rm A}_{2l}=2a^2 + 8ab + 8b^2$;  
$\Lambda^{\rm A}_{3l}=2a^2 + 2ab - 4b^2$;        
$\Lambda^{\rm A}_{4l}=0$. \\            
The parameters for transverse acoustic phonons are:
$\Lambda^{\rm A}_{1t}=\Lambda^{\rm A}_{2t}=\Lambda^{\rm A}_{3t}=0$; 
$\Lambda^{\rm A}_{4t}=\frac{3d^2}{2}$. \\
There are 4 $B_n$ terms:
\begin{equation}		
B_1=S_{11}S^*_{11} + S_{22}S^*_{22}
+   S_{11}S^*_{22} + S_{22}S^*_{11} 
\end{equation}
\begin{equation}	
B_2=S_{33}S^*_{33}
\end{equation}
\begin{equation}	
B_3=S_{11}S^*_{33} + S_{33}S^*_{11}
+   S_{22}S^*_{33} + S_{33}S^*_{22} 
\end{equation}
\begin{equation}	
B_4=\sum_{i}
  S_{i3}S^*_{i3} + S_{i3}S^*_{3i}
+ S_{3i}S^*_{i3} + S_{3i}S^*_{3i}\,.
\end{equation}
The parameters for longitudinal acoustic phonons are:
$\Lambda^{\rm B}_{1l}=2a^2 - 4ab + 2b^2$;
$\Lambda^{\rm B}_{2l}=2a^2 + 8ab + 8b^2$;               
$\Lambda^{\rm B}_{3l}=2a^2 + 2ab - 4b^2$; 
$\Lambda^{\rm B}_{4l}=0$. \\
The parameters for transverse acoustic phonons are:
$\Lambda^{\rm B}_{1t}=\Lambda^{\rm B}_{2t}=\Lambda^{\rm B}_{3t}=0$;
$\Lambda^{\rm B}_{4t}=\frac{3d^2}{2}$. 

\subsection{$[110]$ wires}

For a $[110]$ wire, $\theta_0 = \pi/2$ and $\varphi_0 = \pi/4$. With indices $i,j,k,l,m,n\in\{1,2\}$, the 8 $A_n$ terms are:
\begin{equation}		
A_1=\sum_{i,j,k} O^{ij}_{kkkk}
\end{equation}
\begin{equation}	
A_2=\sum_{i,j} O^{ij}_{3333}
\end{equation}
\begin{equation}	
A_3=\sum_{i,j, k \neq l} O^{ij}_{kkll}
\end{equation}
\begin{equation}	
A_4=\sum_{i,j,k} O^{ij}_{kk33} + O^{ij}_{33kk}
\end{equation}
\begin{equation}	
A_5=\sum_{i,j,k,l \neq m} O^{ij}_{kklm} + O^{ij}_{lmkk}
\end{equation}
\begin{equation}	
A_6=\sum_{i,j,k \neq l} O^{ij}_{33kl} + O^{ij}_{kl33}
\end{equation}
\begin{equation}	
A_7=\sum_{i,j,k \neq l, m \neq n} O^{ij}_{klmn}
\end{equation}
\begin{equation}	
A_8=\sum_{i,j,k,l} O^{ij}_{k3l3} + O^{ij}_{k33l} + O^{ij}_{3kl3} + O^{ij}_{3k3l}\,.
\end{equation}
The parameters for longitudinal acoustic phonons are:
$\Lambda^{\rm A}_{1l}=\frac{4a^2 + 4ab + b^2}{4}$;    
$\Lambda^{\rm A}_{2l}=a^2 - 2ab + b^2$;              
$\Lambda^{\rm A}_{3l}=\frac{4a^2 + 4ab + b^2}{4}$; 
$\Lambda^{\rm A}_{4l}=\frac{2a^2 -  ab - b^2}{2}$;   
$\Lambda^{\rm A}_{5l}=\sqrt{3}d \frac{2a + b}{4}$;
$\Lambda^{\rm A}_{6l}=\sqrt{3}d \frac{a - b}{2}$;   
$\Lambda^{\rm A}_{7l}=\frac{3d^2}{4}$;                
$\Lambda^{\rm A}_{8l}=0$. \\                       
The parameters for transverse acoustic phonons are:
$\Lambda^{\rm A}_{1t}=\frac{9b^2}{4}$;                                       
$\Lambda^{\rm A}_{3t}=-\frac{9b^2}{4}$;                                     
$\Lambda^{\rm A}_{8t}=\frac{3d^2}{8}$;                                      
$\Lambda^{\rm A}_{2t}=\Lambda^{\rm A}_{4t}=\Lambda^{\rm A}_{5t}=\Lambda^{\rm A}_{6t}=\Lambda^{\rm A}_{7t}=0$. \\
There are 8 $B_n$ terms:
\begin{equation}		
B_1=S_{11}S^*_{11} + S_{22}S^*_{22}
\end{equation}
\begin{equation}	
B_2=S_{33}S^*_{33}
\end{equation}
\begin{equation}	
B_3=S_{11}S^*_{22} + S_{22}S^*_{11}
\end{equation}
\begin{equation}	
B_4=S_{11}S^*_{33} + S_{33}S^*_{11}
+   S_{22}S^*_{33} + S_{33}S^*_{22} 
\end{equation}
\begin{equation}	
B_5=\sum_{i \neq j} S_{ii}S^*_{ij} + S_{ii}S^*_{ji}
+                   S_{ij}S^*_{ii} + S_{ji}S^*_{ii} 
\end{equation}
\begin{equation}	
B_6=S_{33}S^*_{12} + S_{33}S^*_{21}
+   S_{12}S^*_{33} + S_{21}S^*_{33} 
\end{equation}
\begin{equation}	
B_7=S_{12}S^*_{12} + S_{12}S^*_{21}
+   S_{21}S^*_{12} + S_{21}S^*_{21} 
\end{equation}
\begin{equation}	
B_8=\sum_{ij} S_{i3}S^*_{j3} + S_{i3}S^*_{3j}
+             S_{3i}S^*_{j3} + S_{3i}S^*_{3j}\,.
\end{equation}
The parameters for longitudinal acoustic phonons are:
$\Lambda^{\rm B}_{1l}=\frac{4a^2 + 4ab + b^2}{2}$;   
$\Lambda^{\rm B}_{2l}=2a^2 - 4ab + 2b^2$;              
$\Lambda^{\rm B}_{3l}=\frac{4a^2 + 4ab + b^2}{2}$;   
$\Lambda^{\rm B}_{4l}=2a^2 -  ab - b^2$,    
$\Lambda^{\rm B}_{5l}=\sqrt{3}d \frac{2a + b}{2}$;   
$\Lambda^{\rm B}_{6l}=\sqrt{3}d (a - b)$;     
$\Lambda^{\rm B}_{7l}=\frac{3d^2}{2}$;                 
$\Lambda^{\rm B}_{8l}=0$. \\                      
The parameters for transverse acoustic phonons are:
$\Lambda^{\rm B}_{1t}=\frac{9b^2}{2}$;                                   
$\Lambda^{\rm B}_{3t}=-\frac{9b^2}{2}$;                                    
$\Lambda^{\rm B}_{8t}=\frac{3d^2}{4}$;                                      
$\Lambda^{\rm B}_{2t}=\Lambda^{\rm A}_{4t}=\Lambda^{\rm A}_{5t}=\Lambda^{\rm A}_{6t}=\Lambda^{\rm A}_{7t}=0$.      

\subsection{$[111]$ wires}

For a $[111]$ wire, $\theta_0=\arccos(\sqrt{3}/3)$ and $\varphi_0=\pi/4$. With indices $i,j,k,l,m,n\in\{1,2,3\}$, the 6 $A_n$ terms are: 
\begin{equation}		
A_1=\sum_{i,j,k} O^{ij}_{kkkk} 
\end{equation}
\begin{equation}	
A_2=\sum_{i,j,k \neq l} O^{ij}_{kkll}
\end{equation}
\begin{equation}	
A_3=\sum_{i,j,k \neq l} O^{ij}_{kkkl} + O^{ij}_{kklk} + O^{ij}_{klkk} + O^{ij}_{lkkk}
\end{equation}
\begin{equation}	
A_4=\sum_{i,j,k \neq l \neq m} O^{ij}_{kklm} + O^{ij}_{lmkk}
\end{equation}
\begin{equation}	
A_5=\sum_{i,j,k \neq l} O^{ij}_{klkl} + O^{ij}_{kllk}
\end{equation}
\begin{equation}	
A_6=\sum_{i,j,k \neq l \neq m} O^{ij}_{klkm} + O^{ij}_{klmk} + O^{ij}_{lkkm} + O^{ij}_{lkmk}\,.
\end{equation}
The parameters for longitudinal acoustic phonons are:
$\Lambda^{\rm A}_{1l}=\frac{2a^2}{3}$;                
$\Lambda^{\rm A}_{2l}=\frac{2a^2}{3}$;              
$\Lambda^{\rm A}_{3l}=\sqrt{3}d \frac{2a}{9}$;        
$\Lambda^{\rm A}_{4l}=\sqrt{3}d \frac{2a}{9}$;        
$\Lambda^{\rm A}_{5l}=\frac{2d^2}{9}$;               
$\Lambda^{\rm A}_{6l}=\frac{2d^2}{9}$. \\             
The parameters for transverse acoustic phonons are:
$\Lambda^{\rm A}_{1t}= \frac{4b^2}{3}$;               
$\Lambda^{\rm A}_{2t}=-\frac{2a^2}{3}$;              
$\Lambda^{\rm A}_{3t}= \sqrt{3}d \frac{b}{9}$;     
$\Lambda^{\rm A}_{4t}=-\sqrt{3}d \frac{2b}{9}$;      
$\Lambda^{\rm A}_{5t}= \frac{d^2}{9}$;               
$\Lambda^{\rm A}_{6t}=-\frac{d^2}{18}$. \\            
There are 6 $B_n$ terms:
\begin{equation}		
B_1=\sum_{i} S_{ii}S^*_{ii}
\end{equation}
\begin{equation}	
B_2=\sum_{i \neq j } S_{ii}S^*_{jj}
\end{equation}
\begin{equation}	
B_3=\sum_{i \neq j } S_{ii}S^*_{ij} + S_{ii}S^*_{ji}
+ S_{ij}S^*_{ii} + S_{ji}S^*_{ii} 
\end{equation}
\begin{equation}	
B_4=\sum_{i \neq j \neq k} S_{ii}S^*_{jk} + S_{jk}S^*_{ii}
\end{equation}
\begin{equation}	
B_5=\sum_{i \neq j} S_{ij}S^*_{ij} + S_{ij}S^*_{ji}
\end{equation}
\begin{equation}	
B_6=\sum_{i \neq j \neq k} S_{ij}S^*_{ik} + S_{ij}S^*_{ki}
+ S_{ji}S^*_{ik} + S_{ji}S^*_{ki}\,.
\end{equation}
The parameters for longitudinal acoustic phonons are:
$\Lambda^{\rm B}_{1l}= 2a^2$;                                     
$\Lambda^{\rm B}_{2l}= 2a^2$;                                  
$\Lambda^{\rm B}_{3l}=\sqrt{3}d \frac{2a}{3}$;                    
$\Lambda^{\rm B}_{4l}=\sqrt{3}d \frac{2a}{3}$;                    
$\Lambda^{\rm B}_{5l}=\frac{2d^2}{3}$;                              
$\Lambda^{\rm B}_{6l}=\frac{2d^2}{3}$. \\                            
The parameters for transverse acoustic phonons are:
$\Lambda^{\rm B}_{1t}=4b^2$;                                     
$\Lambda^{\rm B}_{2t}=-2b^2$;                                  
$\Lambda^{\rm B}_{3t}= \sqrt{3}d \frac{b}{3}$;                    
$\Lambda^{\rm B}_{4t}=-\sqrt{3}d \frac{2b}{3}$;                   
$\Lambda^{\rm B}_{5t}=\frac{d^2}{3}$;                            
$\Lambda^{\rm B}_{6t}=-\frac{d^2}{6}$.                             

\end{document}